\documentclass[a4paper,11pt]{article}
\pdfoutput=1

\usepackage{jheppub}
\usepackage{graphicx}
\usepackage{amsmath}
\usepackage{tabularx}
\usepackage{lscape}

\newcommand{\eq}[1]{eq.~\eqref{eq:#1}}
\newcommand{\eqs}[2]{eqs.~\eqref{eq:#1} and \eqref{eq:#2}}
\newcommand{\eqst}[2]{eqs.~\eqref{eq:#1} - \eqref{eq:#2}}
\newcommand{\eqss}[3]{eqs.~\eqref{eq:#1}, \eqref{eq:#2}, and \eqref{eq:#3}}

\newcommand{\subsec}[1]{section~\ref{subsec:#1}}

\newcommand{\app}[1]{appendix~\ref{sec:#1}}
\newcommand{\fig}[1]{figure~\ref{fig:#1}}

\newcommand{\mycites}[1]{refs.~\cite{#1}}
\newcommand{\mycite}[1]{ref.~\cite{#1}}
\newcommand{\tab}[1]{table~\ref{tab:#1}}

\newcommand{\abs}[1]{\lvert#1\rvert}

\newcommand{\ord}[1]{{\mathcal O}(#1)}

\newcommand{\ORD}[1]{{\mathcal O}\biggl(#1\biggr)}

\newcommand{\convolve}[2]{\left(#1 \otimes #2 \right)(t^\cut)}

\newcommand{\df}{\mathrm{d}}

\newcommand{\cM}{\mathcal{M}}
\newcommand{\cI}{\mathcal{I}}

\newcommand{\e}{\epsilon}

\newcommand{\bn}{\bar{n}}

\newcommand{\nn}{\nonumber}

\newcommand{\lqcd}{\Lambda_\mathrm{QCD}}

\newcommand{\Tau}{\mathcal{T}}
\newcommand{\cut}{\mathrm{cut}}
\newcommand{\bare}{\mathrm{bare}}
\newcommand{\one}{{(1)}}
\newcommand{\two}{{(2)}}
\newcommand{\indep}{\mathrm{indep}}
\newcommand{\inpol}{\mathrm{in}}
\newcommand{\outpol}{\mathrm{out}}
\newcommand{\jet}{\mathrm{jet}}

\newcommand{\msb}{{\overline{\rm MS}}}

\allowdisplaybreaks[2]

\title{Two-Loop Beam and Soft Functions for Rapidity-Dependent Jet Vetoes}

\author[a]{Shireen Gangal,}
\author[b]{Jonathan R.~Gaunt,}
\author[c]{Maximilian Stahlhofen,}
\author[a]{Frank J.~Tackmann}

\affiliation[a]{Theory Group, Deutsches Elektronen-Synchrotron (DESY), D-22607
Hamburg, Germany}
\affiliation[b]{Nikhef Theory Group and VU University Amsterdam, De Boelelaan
1081, NL-1081 HV Amsterdam, the Netherlands}
\affiliation[c]{PRISMA Cluster of Excellence, Institute of Physics, Johannes Gutenberg University, Staudingerweg 7, D-55128 Mainz, Germany}

\emailAdd{shireen.gangal@desy.de}
\emailAdd{jgaunt@nikhef.nl}
\emailAdd{mastahlh@uni-mainz.de}
\emailAdd{frank.tackmann@desy.de}

\abstract{Jet vetoes play an important role in many analyses at the LHC. Traditionally, jet vetoes have been imposed using a restriction
on the transverse momentum $p_{Tj}$ of jets. Alternatively, one can also consider jet observables for which $p_{Tj}$ is weighted by a smooth function of the jet rapidity $y_j$ that vanishes as $|y_j| \to \infty$. Such observables are useful as they provide a natural way to impose a tight veto on central jets but a looser one at forward rapidities. We consider two such rapidity-dependent jet veto observables, $\Tau_{Bj}$ and $\Tau_{Cj}$, and compute the required beam and dijet soft functions for the jet-vetoed color-singlet production cross section at two loops. At this order, clustering effects from the jet algorithm become important.
The dominant contributions are computed fully analytically while corrections that are subleading in the limit of small jet radii are expressed in terms of finite numerical integrals.
Our results enable the full NNLL$'$ resummation and are an important step towards N$^3$LL resummation for cross sections with a $\Tau_{Bj}$ or $\Tau_{Cj}$ jet veto.}


\preprint{\vbox{
\hbox{MITP/16-078}
\hbox{DESY 16-154}
\hbox{NIKHEF 2016-032}
\hbox{August 05, 2016}
}}

\begin{document}

\maketitle

\section{Introduction}
\label{sec:intro}

Jet vetoes find frequent application at the LHC, e.g. in Higgs property measurements as well as in searches for physics beyond the Standard Model. They are used to cut away backgrounds, and more generally to classify the data into exclusive categories, or `jet bins', based on the number of jets in the final state, in order to increase the signal sensitivity. When a generic jet veto observable $\Tau$ is constrained to be much smaller than the hard scale of the process $Q$, large logarithms of $\Tau/Q$ appear in the perturbative expansion of the jet-vetoed cross section, and should be resummed to obtain precise predictions.

The default jet variable by which jets are currently classified and vetoed is the transverse momentum $p_{Tj}$ of a jet.
However, there are some drawbacks to using this variable. In harsh pile-up conditions, it can be difficult to identify (and veto) low-$p_T$ jets in the forward region (beyond $|\eta| \gtrsim 2.5$) when a large part or all of the jet lies in a detector region where no tracking information is available. One way to get around this problem is by introducing a hard cut on the jet (pseudo)rapidity and only consider jets at $|\eta_j| < \eta^\cut$. However, such a cut also changes the logarithmic structure~\cite{Tackmann:2012bt}, and none of the extant resummations of $p_{Tj}$ take account of such a rapidity cut. Another way one might try to avoid the problem is to raise the cut on $p_{Tj}$, but then one loses the potential benefits of a tight jet veto (such as its utility to identify the initial state of heavy resonances~\cite{Ebert:2016idf}).

An alternative approach is to consider a generalized jet-veto variable~\cite{Gangal:2014qda}
\begin{equation}
\Tau_{fj} = p_{Tj} f(y_j)
\end{equation}
that is smoothly dependent on the jet rapidity $y_j$, where the function $f(y)$ is decreasing for increasing $\abs{y}$ such that limiting $\Tau_{fj}$ to small values tightly constrains central jets, but only loosely constrains the forward ones.
In addition to the above practical considerations, given the importance of jet binning
it is highly desirable to have several different options for performing jet vetoes, both experimentally and theoretically. This avoids having to rely exclusively on $p_{Tj}$, and provides important complementary information on the pattern of additional jets produced, for example in Higgs production~\cite{Aad:2014lwa}.

In this paper, we consider the two representative jet observables%
\footnote{In \mycite{Gangal:2014qda}, these definitions were denoted with an additional subscript ``cm'' to distinguish them from the corresponding boost-invariant versions, for which $y_j$ is replaced by $y_j - Y$, where $Y$ is some reference rapidity (e.g. that of the color singlet in color-singlet production).
For our purpose of calculating the relevant soft and beam functions this distinction is irrelevant as the factorization theorems for the corresponding jet-vetoed cross sections only differ by the arguments of the beam functions. We will therefore drop the cm subscript for simplicity of notation in this paper.}
\begin{equation} \label{eq:TauBTauCdefs}
\Tau_{Bj} = m_{Tj} e^{-|y_j|} = \min\{ p_j^+, p_j^- \}
\,, \qquad
\Tau_{Cj} = \frac{m_{Tj}}{2\cosh y_j} = \frac{p_j^+ \, p_j^-}{p_j^+ + p_j^-}
\,,\end{equation}
We have introduced light-cone coordinates, where an arbitrary four-vector $q^\mu$ is decomposed as $q^\mu = q^- n^\mu/2 + q^+ \bn^\mu/2 + q_\perp^\mu$ with $n^\mu$ and $\bn^\mu$ being light-like vectors ($n^2 = \bn ^2 = 0$, $n \cdot \bn = 2$) along the beam directions. Hence, $p_j^\pm = E_j \mp p_j^z$ and so
$m_{Tj}^2 = p_j^+ p_j^- = m_j^2 + p_{Tj}^2$ and $y_j = \ln(p_j^-/p_j^+)/2$.
$\Tau_{Bj}$ gives the plus (minus) momentum of the jet $j$ if the jet lies in the right (left) hemisphere with $p_j^{-(+)} > p^{+(-)}_j$. That is, it has the same rapidity weighting as the global beam thrust hadronic event shape~\cite{Stewart:2009yx}. $\Tau_{Cj}$ has the same rapidity weighting as the $C$-parameter event-shape for $e^+ e^- \to$ dijet processes. It becomes equal to $\Tau_{Bj}$ at forward rapidities and approaches $p_{Tj}/2$ at central rapidities.
The $\Tau_{Cj}$ spectrum has been measured in Higgs production by ATLAS~\cite{Aad:2014lwa}.

In \mycites{Tackmann:2012bt, Gangal:2014qda}, the factorization of the color-singlet production cross sections with a $\Tau_{Bj}$ or $\Tau_{Cj}$ veto, involving jet-dependent beam and soft functions, was formulated within soft collinear effective theory (SCET)~\cite{Bauer:2000ew, Bauer:2000yr, Bauer:2001ct, Bauer:2001yt, Bauer:2002nz, Beneke:2002ph}, and the resummation of the two observables was performed to the NLL$'+$NLO order~\cite{Gangal:2014qda}.

The resummation for the corresponding global beam thrust in color-singlet production is currently known to NNLL$'$~\cite{Stewart:2009yx, Berger:2010xi} (with the results of \mycites{Gaunt:2014xga, Gaunt:2014cfa}). The resummation for a jet-algorithm dependent veto using $p_{Tj}$ is known up to NNLL$'$~\cite{Banfi:2012yh, Becher:2012qa, Tackmann:2012bt, Banfi:2012jm, Becher:2013xia, Stewart:2013faa} and has been applied to a number of color-singlet processes~\cite{Shao:2013uba, Li:2014ria, Moult:2014pja, Jaiswal:2014yba, Becher:2014aya, Wang:2015mvz, Banfi:2015pju, Tackmann:2016jyb, Ebert:2016idf}.

The purpose of the present paper is to provide the full two-loop corrections to the $\Tau_{Bj}$-dependent and $\Tau_{Cj}$-dependent beam and soft functions that are required to bring the resummation for $\Tau_{Bj}$ and $\Tau_{Cj}$ to the NNLL$'$ level. These corrections also provide the necessary fixed-order boundary conditions for the N$^3$LL resummation, with the remaining missing ingredients being the three-loop clustering correction to the noncusp and the four-loop correction to the cusp anomalous dimensions.

We compute, partly analytically and partly numerically, the beam and dijet soft functions for both rapidity-dependent jet veto observables in \eq{TauBTauCdefs}, for all possible color and parton channels. To be precise, for the beam functions we compute the two-loop perturbative matching coefficients between the beam functions and the standard parton distribution functions (PDFs). This is the first explicit calculation of the full set of two-loop singular matching corrections for a jet-algorithm dependent jet veto. This includes the complete set of corrections arising from the clustering of two independent emissions.
(In \mycite{Stewart:2013faa}, the full two-loop soft function for $p_{Tj}$ was calculated, while the two-loop beam functions required for the NNLL$'$ resummation was extracted numerically from fixed-order codes.)
Our results for the two-loop beam functions can also be used in computations of $N$-jet cross sections with a veto on further jets being imposed via $\Tau_{Bj}$ or $\Tau_{Cj}$.

This paper is structured as follows: In section \ref{sec:calc}, we define precisely the soft and beam functions for $\Tau_{Bj}$ and $\Tau_{Cj}$ vetoes, and give an outline of their calculation at two loops. Additional technical details are relegated to the appendices.
In section \ref{sec:results}, we give the obtained results for the two-loop beam and soft functions, and we conclude in section \ref{sec:conclusion}.

\section{Calculation}
\label{sec:calc}

The measurement function for a jet veto using a generic variable $\Tau_{fj}$ is given by
\begin{equation}  \label{eq:Mjetfull}
\mathcal{M}_f^\jet(\Tau^\cut) = \prod_{j \in J(R)} \theta(\Tau_{fj} < \Tau^\cut)
\,.\end{equation}
This constraints all jets to have $\Tau_{fj} < \Tau^\cut$ and thus vetoes any jets with $\Tau_{fj} > \Tau^\cut$.
The jets $J(R)$ in \eq{Mjet} are identified with a specific jet algorithm with jet radius $R$.
At NNLO, we can have at most two emissions, and the precise form of the jet algorithm is irrelevant.
Our results are valid for any jet algorithm that clusters the emissions together if they are closer in
$\Delta R^2 = \Delta \phi^2 + \Delta y^2$ than $R^2$, where $\Delta y$ is the separation in rapidity, $\Delta \phi$ is the separation in azimuthal angle and $R$ is the jet radius. In experimental analyses at the LHC typically $R = 0.4$ or $0.5$.
For two real emissions in the final state with momenta $k_1$ and $k_2$, \eq{Mjetfull} reduces to
\begin{equation}  \label{eq:Mjet}
\mathcal{M}_f^\jet(\Tau^\cut)
= \theta(\Delta R < R)\, \theta (\Tau_{fj} < \Tau^\cut)
+ \theta(\Delta R > R)\, \theta (\Tau_{f1} < \Tau^\cut)\, \theta(\Tau_{f2} < \Tau^\cut)
\,.\end{equation}
That is, for $\Delta R < R$ the two emissions are clustered into a jet and $\Tau_{fj}$ is computed from the sum $p_j = k_1 + k_2$. For $\Delta R > R$, each emissions forms its own jet and $\Tau_{f1}$ and $\Tau_{f2}$ are computed with $p_j = k_1$ and $p_j = k_2$, respectively.

The measurement function $\mathcal{M}_f^\jet(\Tau^\cut)$ is inserted into the usual SCET operator matrix elements defining the beam and soft functions. In this case, the jets $J(R)$ are obtained purely from the collinear or soft radiation within each sector. For details we refer to \mycites{Tackmann:2012bt, Gangal:2014qda}. The practical implementation of such a measurement in beam and soft function calculations is discussed below.

The factorization of the cross section with a $\Tau_{Bj}$ or $\Tau_{Cj}$ veto in \mycites{Tackmann:2012bt, Gangal:2014qda} is strictly speaking valid only to lowest order in an expansion in $R$. The possibility of clustering independent soft and collinear emissions into the same jet breaks the soft-collinear factorization of the measurement function with the corresponding corrections starting at $\ord{R^2}$~\cite{Tackmann:2012bt}. Since this only affects the measurement itself but not the soft-collinear factorization of the amplitudes and SCET Lagrangian, these soft-collinear clustering corrections can be computed in the effective theory and are included in our results.
We separate out the corresponding contributions in the two-loop beam and soft functions that are associated with the clustering of independent emissions. They are denoted with the subscript `indep' and together with the corrections from soft-collinear clustering reproduce the two-loop clustering behaviour of independent emissions in full QCD (in the singular limit). In section \ref{sec:results}, we give two prescriptions as to how this collection of terms can be treated in the NNLL$'$ resummation.

In addition, at $\ord{R^2}$ (potentially) factorization breaking effects due to Glauber interactions can play a role~\cite{Gaunt:2014ska}. At the perturbative level, they first appear in a nonlogarithmic $\ord{\alpha_s^4}$ diagram, implying that the factorization breaking effects first appear at the N$^4$LL order~\cite{Zeng:2015iba, Rothstein:2016bsq}. They will not be discussed further here.

Our calculation of the beam and soft functions is organized in an expansion in $R$ as well.
We will give terms in this expansion up to orders high enough for all practical purposes.
This expansion only involves even powers of $R$ (up to few exceptional terms at lower orders).
We find the $R^2$ expansion to converge very quickly, suggesting that the relevant
expansion parameter is $(R/R_0)^2$ with $R_0 \simeq 2$. (Similar observations have been made recently also in
other contexts involving small-$R$ expansions, see e.g.~\cite{Dasgupta:2016bnd, Kolodrubetz:2016dzb}.)
As pointed out in \mycite{Tackmann:2012bt}, in the small-$R$ limit one should also consider resumming
the corresponding logarithms $\ln R$ appearing in the jet-vetoed cross section. The dominant contribution
beyond $\ord{\alpha_s^2}$ was obtained in \mycite{Alioli:2013hba}. Their
resummation at the LL was obtained in \mycites{Dasgupta:2014yra, Banfi:2015pju}, and very recently methods have been developed~\cite{Kang:2016mcy, Dai:2016hzf} that could allow one to systematically carry out their resummation to higher orders.

To perform the computation of the jet-algorithm dependent soft and beam functions we follow the same strategy used in \mycites{Tackmann:2012bt, Stewart:2013faa, Banfi:2012yh} by computing the difference to a reference soft or beam function defined with a global (jet-algorithm independent) measurement, with the reference functions having been computed elsewhere in the literature. A key property of the global reference measurements we use is that they coincide with the jet-dependent measurements for the case of one real emission. Then when we compute the differences, we only need to consider the double-real emission amplitudes.
Since the measurement functions are different for $\Tau_{Bj}$ and $\Tau_{Cj}$, we must perform a separate computation of the soft function for the two observables. Since $\Tau_{Cj}$ is equal to $\Tau_{Bj}$ at forward rapidities, the beam function is in fact the same for both observables and there is only computation to be performed.

\subsection{Soft function}
\label{subsec:soft}

We discuss explicitly the calculation of the quark soft function, i.e., where the two partons initiating the hard process are quarks. The results for the two-loop gluon soft function are obtained in the usual way by replacing $C_F \to C_A$ due to Casimir scaling.
Following \mycite{Tackmann:2012bt}, the calculation is done in a different way for the `uncorrelated' $C_F^2$ and the `correlated' $C_F C_A$ and $C_F T_F n_f$ color channels.
The notion of correlated and uncorrelated soft emissions is closely connected to the notion of webs in the context of non-Abelian exponentiation of eikonal matrix elements~\cite{Gatheral:1983cz, Frenkel:1984pz}.
Without the potentially exponentiation breaking measurement operator the soft amplitude factorizes into sums of products of webs. For example in our two-loop calculation the $C_F^2$ part of the double-real soft emission amplitude can be written as the product of two identical one-loop amplitudes (webs) proportional to $C_F$. We therefore regard the two emissions in the $C_F^2$ channel as independent or uncorrelated. In contrast, the $C_F C_A$ and $C_F T_F n_f$ parts of the total soft amplitude are nonfactorizable two-loop webs, and we refer to the corresponding emissions as correlated.
The method of calculation for the correlated and uncorrelated channels is described in the following.

\subsubsection[\texorpdfstring{$C_F^2$}{CF CF} piece]{\boldmath $C_F^2$ piece}
\label{subsubsec:softuncorr}

For the uncorrelated $C_F^2$ part of the bare soft function, we conveniently compute the two-loop correction relative to the expectation from non-Abelian exponentiation for the unclustered case, i.e. half of the one-loop soft function squared. This difference is directly the two-loop clustering correction for independent emissions, so we denote it by $\Delta S_{f,\indep}^{\bare\two}(\Tau^\cut,R)$.
The total $C_F^2$ contribution to the bare soft function is then
\begin{equation} \label{eq:SuncorrBmaster}
S_{f}^{\bare(2,C_F^2)}(\Tau^\cut,R)
= \frac{1}{2} \Bigl[ S^{\bare\one}_{f}(\Tau^\cut) \Bigr]^2 + \Delta S_{f,\indep}^{\bare\two}(\Tau^\cut,R)\,.
\end{equation}
The clustering correction $\Delta S_{f,\indep}$ starts at order $R^2$, as discussed earlier, i.e., non-Abelian exponentiation for the jet-veto soft function works up to terms of order $R^2$.

The first term in \eq{SuncorrBmaster} corresponds to using a reference measurement $\theta(\Tau_{f1} < \Tau^\cut)\,\theta(\Tau_{f2} < \Tau^\cut)$, which separately restricts each emissions irrespective of their separation. The difference to \eq{Mjet} then gives the measurement function $\Delta \mathcal{M}_{f,\indep}$ that corresponds to $\Delta S_{f,\indep}^\two$,
\begin{equation} \label{eq:uncorrdelM}
\Delta \mathcal{M}_{f,\indep}
= \theta(\Delta R < R) \Bigl[ \theta(\Tau_{fj} < \Tau^\cut) - \theta(\Tau_{f1} < \Tau^\cut)\,\theta(\Tau_{f2} < \Tau^\cut) \Bigr]
\,,\end{equation}
where $\Tau_{Bi}= \min\{ k_i^+, k_i^-\}$ and  $\Tau_{Ci}= k_i^+ k_i^-/(k_i^+ + k_i^-)$ for the single parton $i=1,2$.

For the $\Tau_{Bj}$ veto, we split the measurement \eq{uncorrdelM} into two pieces:
\begin{align}
\Delta \mathcal{M}_{B,\indep}
&= \Delta \mathcal{M}_{B,\mathrm{indep,1}} + \Delta \mathcal{M}_{B,\mathrm{indep,2}}
\,,\\[1ex]
\label{eq:DeltaMBuncorr1}
\Delta \mathcal{M}_{B,\indep,1}
&= \theta(\Delta R < R) \Bigl[ \theta(k_1^+ + k_2^+ < \Tau^\cut) - \theta(k_1^+<\Tau^\cut) \theta(k_2^+<\Tau^\cut) \Bigr] 2\,\theta(y_t > 0) \,, \\
\label{eq:DeltaMBuncorr2}
\Delta \mathcal{M}_{B,\indep,2}
&= \theta(\Delta R < R) \Bigl\{ \Bigl[ \theta(k_1^+ + k_2^+ < \Tau^\cut) - \theta(k_1^+<\Tau^\cut) \theta(k_2^+<\Tau^\cut) \Bigr]
\nn \\ &\qquad\qquad\qquad\quad \times
2 \bigl[ \theta(y_1>0)\,\theta(y_2>0) - \theta(y_t>0) \bigr]
\,, \nn \\ & \quad
+ \Bigl[ \theta(k_1^+ + k_2^+ < \Tau^\cut) - \theta(k_1^+<\Tau^\cut) \theta(k_2^-<\Tau^\cut) \Bigr]
\nn \\ &\qquad \times
4 \theta(y_1>0)\, \theta(y_2<0)\, \theta(y_j>0)  \Bigr\}
\,.\end{align}
The jet and total rapidities are defined by
\begin{align}
y_j &= \frac{1}{2} \ln  \frac{k_1^- + k_2^-}{k_1^+ + k_2^+}
\, \qquad \text{and} \qquad
y_t = \frac12 (y_1 + y_2) = \frac14 \Bigl( \ln \frac{k_1^-}{k_1^+} + \ln \frac{k_2^-}{k_2^+}  \Bigr)
\,.\end{align}

For the $\Tau_{Cj}$ veto, we split \eq{uncorrdelM} as follows:
\begin{align}
\Delta \mathcal{M}_{C,\indep}
&= \Delta \mathcal{M}_{C,\mathrm{indep,1}} + \Delta \mathcal{M}_{C,\mathrm{indep,2}} \,,
\\[1ex]
\label{eq:DeltaMCuncorr1}
\Delta \mathcal{M}_{C,\indep,1}
&= \theta(\Delta R < R)  \Bigl[ \theta(\Tau_{C1} + \Tau_{C2} < \Tau^\cut) - \theta(\Tau_{C1} <\Tau^\cut) \theta(\Tau_{C2} <\Tau^\cut) \Bigr]
\,, \\
\label{eq:DeltaMCuncorr2}
\Delta \mathcal{M}_{C,\indep,2}
&= \theta(\Delta R < R) \Bigl[ \theta(\Tau_{Cj} < \Tau^\cut) - \theta(\Tau_{C1}+\Tau_{C2} < \Tau^{\cut}) \Bigr]
\,.\end{align}
In both cases, the contribution to $\Delta S$ associated with the first part of the measurement, $\Delta \mathcal{M}_{\rm{f,indep,1}}$, starts to contribute at order $R^2$ and contains a $1/\epsilon$ divergent piece (using dimensional regularization with $d=4-2\epsilon$), and thus produces a single logarithm of $\Tau^\cut$ at $\ord{R^2}$.
The second part is found to start at order $R^4$ and is finite in four dimensions.%
\footnote{For this reason it could be neglected in \mycite{Tackmann:2012bt}. Naive geometrical considerations suggest that these contributions might already start at $\ord{R^3}$, at least for $\Tau_B$. Indeed we find that each of the two terms in \eq{DeltaMBuncorr2} produce contributions of $\ord{R^3}$, which however cancel in the total result.}
Both contributions can be obtained analytically order by order in the $R^2$ expansion.
This calculation is performed in \app{ressoftuncorr}, and we give the results up to $\ord{R^8}$ in \subsec{resultsindepem}.

The jet veto soft function is renormalized multiplicatively~\cite{Tackmann:2012bt, Gangal:2014qda}. To convert the bare results of the above computation to the renormalized $S^{\two}$, we expand the relation $S^{\bare} = Z \times S$ to second order,
\begin{equation} \label{eq:SuncorrB2R}
S^{\bare\two} = Z^{\two} + Z^{\one}S^{\one} + S^{\two}
\,.\end{equation}
For the case of the uncorrelated soft function contribution we can directly use \eq{SuncorrBmaster} together with the one-loop relation $S^{\bare\one} = Z^\one + S^\one$ and find
\begin{align} 
\frac{1}{2}\Bigl( Z^{\one} + S^{\one}\Bigr)^2 + \Delta S_{\rm indep}^{\bare\two}
&= Z^{(2,C_F^2)} + Z^{\one}S^{\one} + S^{(2,C_F^2)}
\,.\end{align}
Hence, we have
\begin{align}
\Delta S_\indep^{\two} &= \Delta S_\indep^{\bare\two} + \frac{1}{2} \big(Z^{\one} \big)^2 - Z^{(2,C_F^2)}
\,,\nn\\
S_f^{(2,C_F^2)} &= \frac{1}{2}(S_f^{\one})^2 + \Delta S_{f,\indep}^{\two}
\,.\end{align}
This shows that the renormalized $S^{\two}$ for the $C_F^2$ channel is equal to the expectation from non-Abelian exponentiation for the unclustered jet veto, $\frac{1}{2}(S^{\one})^2$, plus the finite part $\Delta S^{\two}_\indep$ of the clustering correction.

\subsubsection[\texorpdfstring{$C_F C_A$ and $C_F T_F n_f$}{CF CA and CF TF nf} pieces]
{\boldmath $C_F C_A$ and $C_F T_F n_f$ pieces}
\label{subsubsec:softcorr}

The correlated $C_F C_A$ and $C_F T_F n_f$ contributions to the soft function are split into three pieces,
\begin{align} \label{eq:StotBC}
S^\two_f(\Tau^\cut,R)
= S^\two_{G,f}(\Tau^\cut) + \Delta S^\two_{\mathrm{base}}(\Tau^\cut,R) + \Delta S^\two_{{\rm rest},f}(\Tau^\cut,R)\,.
\end{align}
Here, $S_{G,f}$ is defined for a (known) global reference measurement.
For $\Tau_{Bj}$, $S_{G,B}(\Tau^\cut)$ is given by the cumulant of the single-differential thrust soft function. It has been computed in \mycites{Kelley:2011ng, Monni:2011gb, Hornig:2011iu}, and the explicit $\Tau_B$-differential expression can be found e.g.\ in \mycite{Gaunt:2015pea}.
$S_{G,C}$ is the cumulant of the $C$-parameter soft function, which is defined e.g.\ in eq.~(28) of \mycite{Hoang:2014wka} and has been obtained numerically in \mycites{Hoang:2014wka, Bell:2015lsf}.
Since the reference measurements are chosen to coincide with the jet measurements for a single real emission, $S^\two_{G,f}$ already contains the correct real-virtual contributions. The measurement function $\Delta \mathcal{M}_{f}$ corresponding to the total difference $\Delta S_{f}^\two = S_f^\two-S_{G,f}^\two$ is thus given by
\begin{align} \label{eq:DeltaMBC}
\Delta \mathcal{M}_{f}
&= \theta(\Delta R < R)\, \theta(\Tau_{fj}<\Tau^\cut) + \theta(\Delta R >R)\, \theta(\Tau_{f1}<\Tau^\cut)\, \theta(\Tau_{f2}<\Tau^\cut)
\nn\\ &\quad
-\theta(\Tau_{f1} + \Tau_{f2} <\Tau^\cut)
\,,\end{align}
where the first line is the full jet measurement for two emissions in \eq{Mjet} and the second line subtracts the reference measurement $\mathcal{M}_{G,f} = \theta(\Tau_{f1} + \Tau_{f2} <\Tau^\cut)$.

The divergences of $S_f$ and $S_{G,f}$ differ by a $1/\epsilon$ term and are the same for $\Tau_{Bj}$ and $\Tau_{Cj}$.
The second quantity $\Delta S^\two_{\mathrm{base}}$ in \eq{StotBC} is designed to capture this divergence and is the same for both $\Tau_{Bj}$ and $\Tau_{Cj}$. The remaining piece $\Delta S^\two_{{\rm rest},f}$ is then a finite correction.
The measurement function for $\Delta S^\two_{\mathrm{base}}$ is defined as
\begin{equation} \label{eq:deltaMref}
\Delta \mathcal{M}_{\mathrm{base}}
= 2 \theta (y_t > 0)\, \theta(\Delta R > R) \Bigl[ \theta(k_1^+\! < \Tau^\cut)\, \theta(k_2^+\! < \Tau^\cut) - \theta(k_1^+ + k_2^+\! < \Tau^\cut) \Bigr]
.\end{equation}
The global reference measurement essentially amounts to always clustering the emissions, and $\Delta \mathcal{M}_{\mathrm{base}}$ corrects this to a constraint on the individual emissions when they are further apart than $R$. The reference measurement thus already captures most of the singularities. The remaining $1/\epsilon$ divergence of $\Delta S_f^\two$
is associated with the limit of large jet rapidity $|y_j| \to \infty$.
The correlated amplitude has support only in a finite range of $\Delta y=y_1-y_2$ around zero.
The limits $|y_j| \to \infty$, $|y_t| \to \infty$ and $y_1,\, y_2 \to \pm \infty$ simultaneously are therefore effectively equivalent. 
In particular, in the latter limit $\Delta \cM_f$ in \eq{DeltaMBC} becomes equal to $\Delta \cM_{\mathrm{base}}$ in \eq{deltaMref}. Thus the divergence of $\Delta S_{f}^\two$ is equal to the one of $\Delta S^\two_{\mathrm{base}}$.

As we will see below, the $1/\epsilon$ divergence in $\Delta S^\two_{\mathrm{base}}$ can be isolated analytically, which makes $\Delta S^\two_{\mathrm{base}}$ much easier to compute than the full difference $\Delta S^\two_f$.
In this sense, $\Delta S^\two_{\mathrm{base}}$ performs a very similar role as a subtraction term in a conventional fixed-order calculation.
In particular, the remainder $\Delta S^\two_{{\rm rest},f} = \Delta S^\two_f - \Delta S^\two_{\mathrm{base}}$ can be computed numerically setting $d=4$ from the start.

The soft function $S^\two_{f}$ contains terms proportional to $\ln R$ and $\ln^2\!R$.
These logarithms are entirely contained in $\Delta S^\two_{\mathrm{base}}$, such that $\Delta S^\two_{{\rm rest},f}$ is finite as $R\to0$, see \app{DeltaSrestCalc}. We compute these logarithms analytically as detailed below, while all remaining contributions are computed numerically. Some details regarding the setup for the numerical calculations are given in \app{DeltaSrestCalc}.

In the remainder of this section we describe explicitly the computation of $\Delta S^\two_{\mathrm{base}}$. We first write it in terms of the momenta $k_1$ and $k_2$ of the two emitted gluons,
\begin{equation} \label{eq:softcluster}
\Delta S^\two_{\mathrm{base}}
= {\tilde \mu}^{4 \e}  \int\!\! \frac{d^d k_1}{(2\pi)^d}\frac{d^d k_2}{(2\pi)^d} \, \mathcal{A}^\two_{\rm corr} (k_1, k_2) \, \Delta \mathcal{M}_{\mathrm{base}}(k_1,k_2) \,C(k_1) \, C(k_2)
\,.\end{equation}
Here $C(k_i)= 2\pi \delta(k_i^2)\theta(k_i^0)$ denotes a cut propagator, $\mathcal{A}^\two_{\rm corr}(k_1,k_2)$ is the amplitude obtained by adding all $C_F C_A$ and $C_F T_F n_f$ terms of the double-real diagrams as given explicitly in appendix B of Ref.~\cite{Hornig:2011iu}, $\Delta \mathcal{M}_{\mathrm{base}}$ is the measurement function in \eq{deltaMref}, and we have included the usual $\msb$ factor with ${\tilde \mu} = \mu \exp\{ \tfrac12 [\gamma_E - \ln(4 \pi)] \} $.

We can rewrite the phase-space integral in terms of light-cone components in $d=4-2\epsilon$ dimensions and arrive at
\begin{align} \label{eq:softbarecluster}
\Delta S^\two_{\mathrm{base}}
&= \frac{4^{-4-\epsilon} \, e^{\gamma_E 2\epsilon} \mu^{4 \e}}{\pi^5 \, \Gamma(1-2 \epsilon )}
 \int_0^\infty dk_1^+ dk_2^+ dk_1^- dk_2^- (k_1^+ k_1^- k_2^+ k_2^-)^{-\epsilon}
\nn\\ & \quad \times
\int_0^\pi d\Delta\phi \, (\sin\Delta\phi)^{-2\epsilon} \, \mathcal{A}^\two_{\rm corr}(k_1, k_2) \, \Delta \mathcal{M}_{\mathrm{base}}(k_1,k_2)
\,.\end{align}
To incorporate the constraint on $\Delta R$ we write $k_1^\pm$ and $k_2^\pm$ in terms of the variables
$\Delta y$, $y_t$, $\Delta\phi$, $\Tau_T$, and $z$, as in Appendix C of \mycite{Tackmann:2012bt},
\begin{align} \label{eq:variablesTauB}
y_1 &= \frac{1}{2}\ln{\frac{k_1^-}{k_1^+}}
\,,\quad
y_2 = \frac{1}{2}\ln{\frac{k_2^-}{k_2^+}}
\,,\quad
y_t = \frac{1}{2}(y_1+y_2)
\,,\quad
\Delta y = y_1 - y_2
\,, \nn\\
z &= \frac{k_1^+}{k_1^+ + k_2^+}
\,,\quad
\Tau_T = k_1^+ + k_2^+
\,,\quad
k_1^+ = z \Tau_T
\,,\quad
k_2^+ = (1-z) \Tau_T
\,,\nn\\
\cos\Delta \phi
&= \frac{k_1^{\perp} \!\cdot\! k_2^{\perp}}{|k_1^\perp||k_2^\perp|}
= \frac{\frac12 (k_1^+ k_2^- + k_2^+ k_1^-) - k_1 \!\cdot\! k_2}{\sqrt{k_1^+ k_1^- k_2^+ k_2^-}}
\,.\end{align}
The measurement function in \eq{deltaMref} can be expressed in these variables as
\begin{align} \label{eq:meas_frominc}
\Delta \mathcal{M}_{\mathrm{base}}
= \theta(y_t > 0) \, \theta(\Delta R   > R)\,
\theta \Bigl[ \mathcal{T}^{\cut} < \mathcal{T}_T < \frac{\mathcal{T}^{\cut}}{\max(z,1-z)} \Bigr]
\,.\end{align}

As mentioned previously, the integration of $\mathcal{A}^\two_{\rm corr}$ with $\Delta \mathcal{M}_{\mathrm{base}}$ gives terms proportional to $\ln^n R$. 
We calculate these terms analytically by expanding the full integrand of the  $\Delta y, \Delta \phi, z, \Tau_T$ integration (including the amplitude and phase 
space factors) for small $\Delta R \sim \Delta y \sim \Delta \phi$, keeping only the lowest order terms of order $\Delta R^{-2}$. The integral may then be performed analytically by using the relations
\begin{align}
\label{eq:masterintegral1}
&\int_{-\infty}^\infty d\Delta y\int_0^\pi d\Delta \phi \frac{\Delta \phi^{-2 \epsilon}}{\Delta \phi^2 + \Delta y^2}
\,\theta(\Delta \phi^2 + \Delta y^2 > R^2)
\nn\\ & \qquad
= - \pi \ln\frac{R}{2\pi}
+ \epsilon\,\pi \Bigl(\frac{\pi^2}{12} + \ln^2\frac{R}{2} - \ln^2{\pi} \Bigr) + \ord{\epsilon^2}
\,, \\
\label{eq:masterintegral2}
&\int_{-\infty}^\infty d\Delta y\int_0^\pi d\Delta \phi \frac{2\Delta y^2 \Delta \phi^{-2 \epsilon}}{(\Delta \phi^2+ \Delta y^2)^2}
\,\theta(\Delta \phi^2 + \Delta y^2  > R^2)
\nn\\ &\qquad
= \frac{\pi}{2} - \pi\ln\frac{R}{2\pi}
+ \epsilon\,\pi \Bigl(\frac{1}{2} + \frac{\pi^2}{12} + \ln^2\frac{R}{2} - \ln\frac{R}{2} - \ln^2{\pi} \Bigr)
+ \ord{\epsilon^2}
\,.\end{align}
To compute the remainder of $\Delta S^\two_{\mathrm{base}}$, we subtract the expanded integrand from the full one and integrate the result. The integrand has a sufficiently simple dependence on $y_t$ and $\Tau_T$ that we can perform the integrations over these variables first analytically. The integral over $y_t$ simply gives a $1/(4\epsilon)$ factor. The result can then be expanded in $\epsilon$ up to order $\epsilon^0$ and integrated numerically over the remaining variables $\Delta y$, $\Delta \phi$, $z$ order-by-order in $\epsilon$.

The $\ln R$ and $\ln^2\!R$ terms in the soft and beam functions all arise from \eqs{masterintegral1}{masterintegral2}. They are remainders of the collinear divergence between particles $1$ and $2$, which is regulated by $R$.
For a $p_{Tj}$ veto there is only a single $\ln R$. The $\ln^2\!R$ terms we find for the $\Tau_{fj}$ veto are related to the different treatment of rapidity divergences compared to the $p_{Tj}$ case. In the latter an additional rapidity regulator is required, whereas in our $\Tau_{fj}$ veto calculation rapidity-type divergences are effectively regulated (along with the usual IR and UV divergences) by dimensional regularization. The $\ln^2\!R$ terms only arise from the $\ord{\epsilon}$ terms in \eqs{masterintegral1}{masterintegral2} and therefore cannot appear in the anomalous dimensions. For the same reason, their coefficient in the fixed-order terms is the same as the coefficient of the $\ln R$ terms in the anomalous dimensions. Consequently, they cancel between the beam and soft functions at fixed order due to RG consistency.%

To convert the bare result into a renormalized one, we again use \eq{SuncorrB2R}. For the correlated channels, the cross term $Z^{\one} S^{\one}$ however vanishes, implying that $S^{\two}$ is just the finite part of $S^{\bare\two}$ here. In particular, we can obtain the $C_F C_A$ and $C_F T_F n_f$ contributions to the renormalized jet-dependent $S_{f}^\two$ exactly as shown in \eq{StotBC}, namely by adding to the corresponding reference $S_{G,f}^\two$ the finite part of $\Delta S^\two_{\mathrm{base}}$ as well as $\Delta S^\two_{{\rm rest},f}$.

\subsection{Beam Function}
\label{subsec:beam}

For the beam function, the calculation is done in the same way for all color and partonic channels.
The jet-veto beam functions $B_i$ can be computed as a convolution of perturbative matching coefficients $\cI_{ij}$ with the standard parton distribution functions (PDFs) $f_j$ as~\cite{Fleming:2006cd, Stewart:2009yx, Stewart:2010qs}
\begin{align} \label{eq:BF_OPE}
B_{i}(t^{\cut}\!,x,R,\mu) &= \sum_j \!\int^1_x \!\! \frac{\df z}{z}\, \cI_{ij}(t^{\cut}\!,z,R,\mu) f_{j}\Bigl(\frac{x}{z},\mu \Bigr)
\biggl[1 + \ORD{\frac{\lqcd^2}{t^{\cut}}} \biggr]
\,.\end{align}
To obtain the matching coefficients at two loops, we compute the two-loop partonic beam functions $B_{ij}$, see \mycites{Stewart:2009yx, Stewart:2010qs, Gaunt:2014xga, Gaunt:2014cfa} for details and related definitions in terms of SCET operator matrix elements.

The partonic jet-dependent $B_{ij}$ are calculated via the difference $\Delta B_{ij}$ between them and the reference beam functions as
\begin{equation} \label{eq:Bref}
 B_{ij}(t^\cut,x,R) = B_{G,ij}(t^\cut,x) + \Delta B_{ij}(t^\cut,x,R)
\,,\end{equation}
where $t^\cut \equiv xp^-\Tau^\cut$ with $p^-$ being the large light-cone momentum of the incoming parton.
The reference functions $B_{G,ij}(t^\cut,x)$ are defined using the global reference measurement $\theta(\Tau < \Tau^\cut)$, where $\Tau$ is the total plus momentum of all real emissions. They are given by the cumulant of the virtuality-dependent beam functions calculated in \mycites{Gaunt:2014xga, Gaunt:2014cfa},
\begin{equation}
B_{G,ij}(t^\cut,x) = \int_0^{t^\cut} \!\df t \, B_{ij} (t,x)
\,.\end{equation}

The measurement function corresponding to $\Delta B_{ij}$ is given by the difference of \eq{Mjet} and the global reference measurement,
\begin{align} \label{eq:clusteringMbeam}
\Delta \mathcal{M}^\jet(\Tau^\cut)
&= \theta(\Delta R <  R)\, \theta(\Tau_j < \Tau^\cut) + \theta(\Delta R > R)\,\theta(\Tau_1 < \Tau^\cut)\, \theta(\Tau_2 < \Tau^\cut)
\nn\\ & \quad
- \theta(\Tau_{1} + \Tau_{2} < \Tau^\cut)
\nn\\
&= \theta(\Delta R > R)\, \Bigl[\theta(k_1^+ < \Tau^{\cut}) \theta(k_2^+ < \Tau^{\cut})- \theta(k_1^+ + k_2^+ < \Tau^{\cut}) \Bigr]
\,.\end{align}
In the second step we used that in the collinear sector we always have $\Tau_i = k_i^+$ and $\Tau_j = k_1^+ + k_2^+$ for both $\Tau_{Bj}$ and $\Tau_{Cj}$.
In addition, we have the usual measurement $\delta$-function that fixes the large light-cone momentum fraction of the parton that enters the hard process to $x$.

Although collinear matrix elements, like the beam functions, do not exponentiate, we nevertheless introduce a notion of `correlated' and `uncorrelated' emissions also for their calculation. We note, however, that this distinction is purely technical and there is no one-to-one correspondence to the different color factors.
It is inspired by the behavior of the amplitude in the limit where (at least) one emission becomes soft.

We start by taking the double-real emission amplitudes $\mathcal{A}$ for the partonic beam function for each parton and color channel,
previously calculated in \mycites{Gaunt:2014xga, Gaunt:2014cfa}. These amplitudes
are gauge invariant and have been calculated in both Feynman and axial (light-cone)
gauge as a cross check. We again denote the momenta of the emitted partons as $k_1$ and $k_2$,
and (without loss of generality) we have symmetrized the amplitudes such that they are symmetric under interchanging
$k_1 \leftrightarrow k_2$. For each color channel we separate $\mathcal{A}$ into two pieces,
\begin{equation} \label{eq:ampsepAB}
\mathcal{A}(k_1,k_2,x) = \mathcal{A}_{A}(k_1,k_2,x) + \mathcal{A}_{B}(k_1,k_2,x)
\,,\end{equation}
where $\mathcal{A}_{A}$ is defined by
\begin{equation} \label{eq:AAterm}
\mathcal{A}_{A}(k_1,k_2,x)
\equiv \frac{\lim_{k_1^- \to 0} \bigl[ k_1^- k_2^- \cdot \mathcal{A}(k_1,k_2,x)|_{k_2^-=(1-x)p^--k_1^-} \bigr]}{k_1^-k_2^-}
\,,\end{equation}
and $\mathcal{A}_B = \mathcal{A} - \mathcal{A}_{A}$ is the remainder. 
Labeling the parton entering the hard process with $i$ and the incoming parton with $j$, the term $\mathcal{A}_{A}$ always has the form
\begin{equation} \label{eq:AAterm1}
\mathcal{A}_{A, ij} = \frac{1}{2} \mathcal{A}_{ij}^{B(1)}(k_2,x) \, \mathcal{A}_{i}^{S(1)} (k_1)
\,,\end{equation}
with
\begin{equation}
 \mathcal{A}_{ij}^{B(1)}(k_2,x) = \frac{2g^2\hat{P}^{(0)}_{ij}(x)(1-x)p^-}{k_2^+k_2^-}
\,, \qquad 
 \mathcal{A}_{i}^{S(1)}(k_1) = \frac{4g^2C_i}{k_1^+k_1^-}
\,.\end{equation}
The color factors are defined as
\begin{equation}
\label{eq:Ci}
C_i = \begin{cases}
         C_F &\text{for}\quad i=q
         \\
         C_A &\text{for}\quad i=g
        \end{cases}
\,,\end{equation}
and the $\hat{P}^{(0)}_{ij}(x)$ are the unregularized one-loop splitting functions
\begin{align} \label{eq:oneloopsplit}
\hat{P}^{(0)}_{q_iq_j}(x) &= C_F\,\theta(x)\, \delta_{ij}\hat{P}_{qq}(x),
\,,\\
\hat{P}^{(0)}_{gg}(x) &= C_A\, \theta(x)\, \hat{P}_{gg}(x)
\,, \\
\hat{P}^{(0)}_{q_ig}(x) = \hat{P}^{(0)}_{\bar{q}_ig}(x) &= T_F\,\theta(x)\, \hat{P}_{qg}(x)
\,, \\
\hat{P}^{(0)}_{gq_i}(x) = \hat{P}^{(0)}_{g\bar{q}_i}(x)  &= C_F\, \theta(x)\, \hat{P}_{gq}(x)
\,,\end{align}
where
\begin{align}\label{eq:basicsplits}
\hat{P}_{qq}(x) &= \frac{1+x^2}{1-x}
\,, \\
\hat{P}_{gg}(x) &= 2\Bigl[ \frac{x}{1-x} + \frac{1-x}{x} + x(1-x) \Bigr]
\,, \\
\hat{P}_{qg}(x) &= x^2 + (1-x)^2
\,, \\
\hat{P}_{gq}(x) &= \frac{1+(1-x)^2}{x}
\,.\end{align}
Note that $\mathcal{A}_{A}$ is nonzero only for certain parton and color channels.

As can be seen from \eq{AAterm}, $\mathcal{A}_A$ is essentially obtained by taking the soft limit for one of the partons. In this limit the amplitude, since it contains a sum over all diagrams, falls apart into a product of a one-loop soft amplitude $\mathcal{A}_{i}^{S(1)}$ and a one-loop collinear amplitude $\mathcal{A}_{ij}^{B(1)}$ by Ward identity arguments. As should be clear from its structure, the term $\mathcal{A}_A$ corresponds to the uncorrelated emission part of the amplitude. This term is the part of the amplitude that when expressed in terms of the $\Delta y, \Delta \phi, z, \Tau_T,y_t$ variables in \eq{variablesTauB}, and when multiplied by the appropriate Jacobian becomes flat in  $\Delta y$, $\Delta \phi$ as well as in $y_t$ in four dimensions. It is also the only part of the amplitude that remains in the zero-bin subtraction terms (see below).

The integration of the term $\mathcal{A}_B$ yields only a $1/\epsilon$ divergence, and is essentially identical in character to the integral of the soft correlated amplitude.
In particular there are no divergences associated with the $z, \Delta y, \Delta \phi$ or $\Tau_T$ integrations, so the same integration variables and techniques as for the correlated soft contributions can be used to integrate this piece. Integrating the $\mathcal{A}_A$ piece yields a much deeper divergence structure, but this piece can be integrated in a straightforward way analytically.

\subsubsection{`Correlated' piece}

To integrate $\mathcal{A}_B$, the $y_t$ integration can be performed using the $\delta$-function for the large minus momentum, $\delta [k_1^- + k_2^- = (1-x)p^-]$, after which the expressions for $k_1^\pm$ and $k_2^\pm$ in terms of the remaining integration variables $z,\Delta y, \Delta \phi, \Tau_T$ are
\begin{align}
 & k_1^- = \frac{e^{2\Delta y} p^- (1-x) z}{( e^{2\Delta y}-1) z+1}\,, && k_2^- = \frac{p^- (1-x) (1-z)}{(e^{2\Delta y}-1) z+1} \,,
\\ \nn
 & k_1^+ = z \Tau_T \,, && k_2^+ = (1-z) \Tau_T
\,.\end{align}

The integrand contains an explicit factor of $(1-x)^{-1-2\epsilon}$ which we pull out and expand in terms of distributions:
\begin{equation} \label{eq:distroexp}
 (1-x)^{-1-2\epsilon} = -\frac{1}{2\epsilon}\delta(1-x) + \mathcal{L}_0(1-x) + \dotsb
\,,\end{equation}
where $\mathcal{L}_0(1-x) \equiv 1/(1-x)_+$ is the standard plus distribution. The remaining part of the integrand is then simply expanded in $\epsilon$ and integrated over the remaining variables. There are no additional divergences associated with these integrals, and they yield a function that is finite in the limit $x \to 1$. Note that for the channels with $i\neq j$, this function in fact vanishes for $x \to 1$, such that for these channels we can replace the right hand side of \eq{distroexp} by $1/(1-x)$. This means that there is only a $1/\epsilon$ piece for the $i=j$ channels proportional to $\delta(1-x)$, as one may anticipate from the fact that the anomalous dimension for the beam function is diagonal in flavour.

The integration of $\mathcal{A}_B$ gives terms proportional to $\ln R$ and $\ln^2\!R$ for some color structures. We note in passing that in axial gauge the $\ln^n\!R$ terms arise from only one diagram topology, namely the `bubble insertion' graph shown in figure 2$f$) and 2$o$) of \mycite{Gaunt:2014xga} for the quark beam function and in figure 1$i$) and 1$o$) of \mycite{Gaunt:2014cfa} for the gluon beam function. This topology is also sketched on the left-hand side of \fig{smallRgraph}.
As for the soft function, we calculate the $\ln^n\! R$ terms analytically by expanding the integrand for small $\Delta R$, and performing the integration using \eqs{masterintegral1}{masterintegral2}. Compact expressions for the required small $\Delta R$ limits of the beam function amplitudes are given in \app{smallRformulae}. 

The remaining contributions from $\mathcal{A}_B$ can then be obtained by subtracting the expanded integrand from the full integrand and integrating this numerically, as was done for the correlated soft function contributions. However, here this must be done for various points in {\em both} the $x$ and $R$ directions, requiring to generate a 2D grid of points.
We did compute such a grid -- however, since such results are not so straightforward to use or present, we also follow an alternative approach, the results of which are given in section \ref{sec:results} below. In this alternative approach we explicitly compute the leading $R$-dependent terms proportional to $R^2$ of the non-$\delta(1-x)$ parts of the beam function analytically. The $R^0$ terms in the non-$\delta(1-x)$ piece are computed numerically, along with the full $R$ dependence of the $\delta(1-x)$ piece. For these pieces we can then give simple one-dimensional functions (in either $x$ or $R$) obtained by interpolating the numerically generated points. The results from this approach are sufficiently accurate for smaller $R$ values. More quantitative statements and a comparison of these approximated results to the exact numerical results are given in section \ref{sec:results}.

To compute the terms proportional to $R^2$, we expand the integrand up to order $\Delta R^0$. We then take the $\theta(\Delta R > R)$ in the measurement \eq{clusteringMbeam} and divide it into two pieces,
\begin{equation} \label{eq:thetaRsplit}
 \theta(\Delta R > R) = 1 - \theta(\Delta R < R)
\,,\end{equation} 
which yields the integral
\begin{equation} \label{eq:leadingRIntegral}
\int\! \df \Phi\, \bigl[\mathcal{I}_{-2} + \mathcal{I}_{0} + \mathcal{O}(\Delta R^2)\bigr] \times  \bigl[1 - \theta(\Delta R   < R) \bigr]
\,.\end{equation}
Here $\int\!\df \Phi$ denotes the phase-space integration over all of the relevant integration variables, in particular over $\Delta \phi$ and $\Delta y$. The $\mathcal{I}_{-2}$ term denotes the part of the integrand of order $\Delta R^{-2}$, whilst $\mathcal{I}_{0}$ is the part of $\ord{\Delta R^0}$. It is clear that the integrations involving the `1-term' in the right factor will give a $R$-independent result, which we can drop if we are only interested in the $R^2$ terms. The integration of the remaining $\mathcal{O}(\Delta R^2)$ terms with $\theta(\Delta R < R)$ gives contributions of $\ord{R^4}$, so we can drop these too.
The integration of the $\mathcal{I}_{-2}$ with the $\theta(\Delta R < R)$ generates the above mentioned $\ln^n R$ plus some $\ord{R^0}$ terms. The $R^2$ contributions we want to isolate thus exclusively come from the integral%
\footnote{This technique can easily be extended to obtain the $R^{2n}$ corrections for arbitrary $n$, and we have indeed obtained $R^4$ expressions for the diagonal channels as discussed below.}
\begin{equation} \label{eq:RsqIntegral}
 - \int\! \df \Phi  \, \mathcal{I}_{0} \, \theta(\Delta R < R)
\,.\end{equation}
We performed this integral analytically for the non-$\delta(1-x)$ parts of each beam function.%
\footnote{Of course the analogous analytical calculation can be done for the $\ord{R^2}$ pieces in the $\delta(1-x)$ terms.
For these we can however perform a 1-parameter fit for the full $R$ dependence with $x=1$, which is accurate enough that analytical results of the $R^2$ contributions are not needed.}

The non-$\delta(1-x)$ piece $\propto R^0$ we obtained by numerically computing the integral in \eq{leadingRIntegral} at $R=0.2$, and then extrapolating to $R=0$ using the leading $R$-dependence just obtained, which works very well for the small $R$ values between $0$ and $0.2$.
We do not directly compute the integral at $R=0$ to avoid numerical instabilities.

\subsubsection{`Uncorrelated' piece}

Next, to integrate the uncorrelated $\mathcal{A}_{A}$ part of the amplitude, our method is as follows.
We first write again $\theta(\Delta R > R) = 1 - \theta(\Delta R < R)$ as in \eq{thetaRsplit}.
The integration of $\mathcal{A}_A$ with the `$1$-term' can be straightforwardly done analytically without a change of variables.
This yields terms constant in $R$ that can be as divergent as $1/\epsilon^2$.
Note that the `$1$-term' here corrects the reference measurement to the fully unclustered case.
The $\theta(\Delta R < R)$ term then is the analogue to the clustering correction for the soft uncorrelated emissions
in \eq{uncorrdelM}.

The integration of $\mathcal{A}_A$ with $\theta(\Delta R < R)$ can be done analytically order-by-order in $R$ after changing variables
to $\Delta y$, $\Delta \phi$, $z$, $\Tau_T$, and like in the soft case yields terms starting at $\mathcal{O}(R^2)$.
For the $qq$ $C_F^2$ and $gg$ $C_A^2$ channels only, this piece yields a divergent
$R^2/\epsilon$ term that is exactly the one anticipated in eq.~(50) of \mycite{Tackmann:2012bt}, but with the opposite sign.

For this piece, the zero-bin subtractions are nonvanishing.
Any of the zero-bin limits, $k_1$ soft, $k_2$ soft, or both soft, picks out only the term $\mathcal{A}_{A}$ from the full amplitude $\mathcal{A}$. The zero-bin computation yields terms of $\mathcal{O}(R^2)$. 
For the $qq$ $C_F^2$ and $gg$ $C_A^2$ channels the total
zero bin contribution gives a divergent $R^2/\epsilon$ term, which is exactly twice the
naive result from the $\theta(\Delta R < R) \times \mathcal{A}_A$ term above.
The full $R^2/\epsilon$ contribution, given by subtracting the zero-bin contribution from the naive result,
then exactly reproduces and confirms eq.~(50) of \mycite{Tackmann:2012bt}.
Note that the zero-bin contribution for the `$1$-term' is scaleless and vanishes, i.e.,
the zero-bin for the fully unclustered contribution is scaleless just like in the inclusive case.
We discuss the full computation of the $\theta(\Delta R < R) \times \mathcal{A}_A$ term and the corresponding zero-bin
terms in more detail in \app{zerobins}.

\subsubsection{Renormalization}
\label{subsec:BRenormalization}

The result of the above calculation yields the difference between the bare jet-veto and reference beam functions, $\Delta B^\bare$.
To convert this to a difference between renormalized beam functions, we must
expand the relation $B^\bare = Z \otimes B$ for the two beam functions to two loops.
Even though $Z^{(1)}$ and $B^\one$ are the same for both ($B_G^\one = B^\one$), the two beam functions renormalize differently, i.e., the nature of the convolution between the counter terms $Z$ and the renormalized beam functions $B$ is different.
In particular, for the jet-veto beam function this convolution is just the simple product.
Taking the difference between the renormalized beam functions according to \eq{Bref} we thus obtain
\begin{align} \label{eq:Bren2}
\Delta B^{\two}_{ik}(t^\cut) &
 =  \Delta B_{ik}^{\bare\two}(t^\cut) -  \Delta Z_i^{\two}(t^\cut) \delta_{ik}
\nn \\ & \quad
- \Bigl[ Z_i^\one(t^\cut) B_{ik}^\one(t^\cut) - \convolve{Z_i^\one}{B_{ik}^\one} \Bigr]
\,,\end{align}
where we define
\begin{equation}
 \convolve{A}{B} = \int_0^{t^\cut} \! \df t \int_0^t \! \df t' \, A(t-t') \, B(t')
\,.\end{equation}
The second term in square brackets in \eq{Bren2} accounts for the different renormalization of the global
reference beam function. 

The result for $\Delta Z^{\two}_i$ can be converted to the corresponding clustering correction for the two-loop
anomalous dimensions, $\Delta \gamma^{\two}_{Bi}$. The relation
needed to achieve this can be obtained by expanding $Z\otimes \gamma_B = - \df Z/\df \ln \mu$
to two-loop order for the two beam functions and taking the difference,
\begin{equation} \label{eq:delG2}
\Delta \gamma_B^\two(t^\cut) = - Z^\one(t^\cut) \times \gamma_B^\one (t^\cut)
 + \convolve{Z^\one}{\gamma_B^\one} - \biggl[ \frac{\df \Delta Z(t^\cut)}{\df \ln \mu} \biggr]^\two
\,.\end{equation}
The result we find for $\Delta \gamma_B^\two$ is precisely as was predicted in \mycite{Tackmann:2012bt}. 
Namely, it is $(-1/2)$ times that of the soft function, plus an additional term related to soft-collinear mixing as required by RG consistency. This is an important check of our calculation. 

Finally, the difference in two-loop renormalized partonic beam functions $\Delta B_{ij}^\two$ can be converted into a difference in two-loop matching coefficients $\Delta \cI^\two_{ij}$ by expanding \eq{BF_OPE} for a partonic incoming state to two loops for the jet-veto and reference beam functions, and taking the difference. 
Since $\Delta \cI_{ij}^{(0)}$ and $\Delta \cI_{ij}^\one$ are both zero, this immediately yields
\begin{align}
\Delta \cI^\two_{ij}(t^\cut, x, R, \mu) = \Delta B^\two_{ij}(t^\cut, x, R, \mu)
\,.\end{align}

The contributions to $\Delta \cI^\two_{ij}$ from the integration of $\mathcal{A}_A$ with the unclustered `$1$ term', as well as the terms in square brackets in \eq{Bren2}, are of $\ord{R^0}$. As mentioned above, their purpose is to correct the running of the beam function to be multiplicative rather than involving convolutions. 
We therefore present these pieces together with the integration of the correlated amplitude $\mathcal{A}_B$ below. On the other hand, the integration of $\mathcal{A}_A$ with $\theta(\Delta R < R)$, together with corresponding the zero-bin terms, are of $\ord{R^2}$. They are in fact the clustering corrections for independent emissions (analogous to $\Delta S^\two_{\rm indep}$), and so we separate these pieces off, and denote them using the subscript `indep' in the results below.

\section{Results}
\label{sec:results}

Here we present the two-loop results for the $\Tau_{fj}$-dependent beam and soft functions, together with the finite part of the two-loop soft-collinear mixing terms discussed in \mycite{Tackmann:2012bt}. For completeness, we also present the corresponding anomalous dimensions of these functions. In certain places in the results we give functions of $R$ obtained by fitting numerically generated points. These points were generated using the GlobalAdaptive NIntegrate routine from {\tt Mathematica}, and have a numerical uncertainty at the sub-per-mille level. 
We checked this by using the Suave routine from the Cuba library~\cite{Hahn:2004fe} to perform the same integrations and confirmed that the results agreed at the sub-per-mille level with the GlobalAdaptive method once a sufficiently large number of integration points were used.
The number of data points we used for the fits was $30$ in each case, spanning the range $R=0.05-1.5$. 
We checked that the fit functions reproduce each point at the sub-per-mille level, and that there are no visible interpolation artifacts (such as `polynomial wiggle') in these fits.
In section \ref{subsec:Bcorr} we give functions of $x$ that have also been fitted. 
The points for these fits have also been generated using {\tt Mathematica} NIntegrate with sub-per-mille precision (and checked using Suave). 
The fitting procedure for these functions is slightly more involved and described at the end of section \ref{subsec:Bcorr}.

Our final results for the renormalized two-loop beam and soft functions to be used in the NNLL$'$ resummation are given by%
\footnote{To simplify the results, we explicitly extract the overall factors of $\alpha_s(\mu)$ here, while they are always included in the bare two-loop expressions.}
\begin{align}
S_f(\Tau^\cut,R, \mu)
&= 1 + \frac{\alpha_s(\mu)}{4\pi}\, S^\one_f(\Tau^\cut, \mu) + \frac{\alpha_s^2(\mu)}{(4\pi)^2}\,S^\two_f(\Tau^\cut,R, \mu)
\,, \nn \\
\cI^{\two}_{ij}(t^\cut,x,R, \mu)
&= \delta_{ij} \delta(1 - x) + \frac{\alpha_s(\mu)}{4\pi}\,\cI_{ij}^\one(t^\cut, x, \mu)
 + \frac{\alpha_s^2(\mu)}{(4\pi)^2}\, \cI^{\two}_{ij}(t^\cut,x,R, \mu)
\,,\end{align}
where the one-loop coefficients do not yet depend on $R$ and are simply given by the cumulants of the corresponding differential functions. The two-loop contributions are written as
\begin{align}
\label{eq:masterS}
S^\two_f(\Tau^\cut,R, \mu)
&= S^{(2,C_i C_A)}_{G,f}(\Tau^\cut, \mu) + S^{(2,C_i T_Fn_f)}_{G,f}(\Tau^\cut, \mu)
+ \frac{1}{2}\Bigl[S_f^{\one}(\Tau^\cut, \mu)\Bigr]^2
\\ \nn & \quad
+ \Delta S_{f}^{(2)}(\Tau^\cut,R, \mu) + \Delta \mathbb{S}_f^{(2)}(\Tau^\cut,R, \mu)
\,,\\
\label{eq:masterI}
\cI^{\two}_{ij}(t^\cut,x,R, \mu)
&= \cI^\two_{G,ij}(t^\cut,x, \mu) + \Delta \cI^\two_{ij} (t^\cut,x,R, \mu) + \Delta \mathbb{I}^\two_{ij} (t^\cut,x,R, \mu)
\,.\end{align}
Here, $\cI^\two_{G,ij}$ is the cumulant of the virtuality-dependent beam function matching coefficients, which can straight-forwardly be obtained from the results of \mycites{Gaunt:2014xga, Gaunt:2014cfa}. For $f=B$ the function $S_{G,f}$ is the cumulant of the renormalized thrust soft function and for $f=C$ it is the cumulant of the renormalized C-parameter soft function.

The results for the corrections due to the jet clustering are split into two parts.
The corrections $\Delta S_{f}^{(2)}$ and $\Delta \cI^\two_{ij}$ are not associated with the clustering of independent emissions. For simplicity we call them `correlated' pieces. 
They are given in section \ref{subsec:resultscorr}. Finally, the $\Delta \mathbb{S}_{f, \indep}^{(2)}$ and $\Delta \mathbb{I}^\two_{ij}$ are the corrections from the clustering of independent emissions, which start at $\ord{R^2}$.
They are given in section \ref{subsec:resultsindepem}, together with the associated two-loop soft-collinear mixing terms. There, we also give two alternative prescriptions for incorporating these terms in the resummation at NNLL$'$.

\subsection{`Correlated' pieces} \label{subsec:resultscorr}

\subsubsection{Soft function} \label{subsec:STcorr}

We find
\begin{equation} \label{eq:softresultTauB}
\Delta S_f^{(2)}(\Tau^{\cut},R, \mu)
= \Delta\gamma_{S\,1}^i(R)\, \ln\frac{\mu}{\Tau^\cut} + \Delta s^i_{2f}(R)
\,,\end{equation}
with the anomalous dimension correction
\begin{align}  \label{eq:dS1f}
\Delta\gamma_{S\,1}^i(R)
&= 4 C_i\, C_A \Bigl[
\Bigl(1 - \frac{8 \pi^2}{3}\Bigr) \ln R - \frac{13}{2}  + 6 \ln2 + 4 \mathcal{F}^{C_A}(R) + 11 \mathcal{F}^{T_F}(R)
\Bigr]
\nn \\ & \quad
+ 4 C_i\, \beta_0 \Bigl[\Bigl(\frac{23}{3} - 8 \ln2\Bigr) \ln R + \frac{13}{6} - 2 \ln2 - 3 \mathcal{F}^{T_F}(R) \Bigr]
\,,\end{align}
and%
\footnote{We thank the authors of \mycites{Bell:2018jvf, Bell:2018vaa, Bell:2018oqa}
for pointing out a typo in a previous version,
where the $C_A \ln^2 R$ coefficient had its sign flipped.
In addition, we corrected a bug in the $R$-independent terms for $\Tau_{Bj}$, which
affected the $c_1$ coefficients for $\Tau_{Bj}$ in \tab{finite_soft_coeff}.}
\begin{align}
\Delta s^i_{2f}(R)
&=  16\,C_i\,\biggl\{C_A\biggl[\frac{-131+12\pi^2+132 \ln{2}}{72} \ln^2{R} + \frac{395-33\pi^2-216 \zeta_3}{54} \ln{R} + f_f^{C_A}(R) \biggr]
\nn \\ & \quad
+ T_F n_f \biggl[\frac{ 23-24\ln{2}}{36} \ln^2{R} + \frac{ -245+ 24\pi^2-36\ln{2}}{108} \ln{R} + f_f^{T_F}(R) \biggr] \biggr\}
\,,\end{align}
where $C_i = C_F$ for the quark soft function and $C_i = C_A$ for the gluon soft function.

The functions $\mathcal{F}$ are fitted from numerically generated points using the form
\begin{equation}\label{eq:fitform_F}
\mathcal{F}(R) = a_1 + a_2 R^2 + a_3 R^4
\,,\end{equation}
and the fitted coefficients $a_i$ are given in \tab{anom_soft_coeff}.
The functions $f$ are different for $\Tau_{Bj}$ and $\Tau_{Cj}$. They are fitted from numerically generated points using the form
\begin{equation} \label{eq:fitform_f}
 f(R) = c_1 + c_2 R^2 + c_3 R^4 + c_4 R^2\ln R + c_5 R^4\ln R
\,,\end{equation}
and the fitted coefficients $c_i$ are given in table \ref{tab:finite_soft_coeff}.

\begin{table}
\centering
\begin{tabular}{c | c c c }
\hline\hline
Channel & $a_1$ & $a_2$ & $a_3$ \\ \hline
$C_A$ & $-1.0670$
  & $0.65238$ & $-0.010291$  \\
$T_F$ & $1.0076$
& $0.019958$ & $-1.0523\cdot 10^{-3}$  \\
\hline\hline
\end{tabular}
\caption{Fit function coefficients $a_i$ for the anomalous dimension fitting functions $\mathcal{F}(R) = a_1 + a_2 R^2 + a_3 R^4$.
\label{tab:anom_soft_coeff}
}
\end{table}

\begin{table}
 \centering
\begin{tabular}{c | c c c c c c}
\hline\hline
Channel & $c_1$  & $c_2$ &  $c_3$ & $c_4$ & $c_5$
\\ \hline
$\Tau_{Bj}$: $C_A$ & $-0.62632$ & $0.077171$ & $0.015511$ & $-0.34184$ & $0.023850$
\\
$\Tau_{Bj}$: $T_F$ & $0.36570$  & $-0.032704$ & $-7.2734\cdot 10^{-4}$ & $-0.010864$ & $1.0880\cdot 10^{-3}$
\\\hline
$\Tau_{Cj}$: $C_A$ & $-0.75296$ & $- 0.19674$ & $ 0.067391 $ & $ - 0.33153$ & $ - 0.0038577 $
\\
$\Tau_{Cj}$: $T_F$ & $0.38286$  & $- 0.062270$ & $ 2.9134\cdot10^{-3} $ &  $- 0.010353$ & $ - 6.9651\cdot10^{-4}$
\\\hline\hline
\end{tabular}
 \caption{Fit function coefficients $c_i$ for the correlated soft function $f(R) = c_1 + c_2 R^2 + c_3 R^4 + c_4 R^2\ln R + c_5 R^4\ln R $.}
 \label{tab:finite_soft_coeff}
\end{table}

\subsubsection{Beam function} \label{subsec:Bcorr}

We decompose the two-loop matching coefficient corrections $\Delta \cI^\two_{ij}$ as follows
\begin{align}
\Delta \cI^\two_{ij}(t^\cut, x, R, \mu)
&= \delta_{ij}\,\frac{\Delta\gamma_{S\,1}^i(R)}{4} \Bigl[\delta(1 - x) \ln\frac{t^\cut}{\mu^2} + \mathcal{L}_0(1-x)\Bigr]
+ \Delta \cI^\two_{ij,\mathrm{run}}(t^\cut, x, \mu)
\nn \\ & \quad
+ 4\Delta I_{ij}^\two(x, R)
\,,\end{align}
where $\Delta \gamma_{S\,1}(R)$ is given in \eq{dS1f}.
The functions $\Delta \cI^\two_{ij,\mathrm{run}}(x)$ are $R^0$ terms and contain the $\mu$ dependent terms that are required to convert the running of the beam function to be multiplicative (see \subsec{BRenormalization}),
\begin{align} 
\Delta \cI^\two_{ij,\mathrm{run}}(t^\cut,x, \mu)
&= 16C_i\, \biggl\{
C_i \delta_{ij} \, \delta(1-x)\,
\Bigl(  \frac{ \pi ^2}{12} \ln ^2 \frac{t^\cut}{\mu^2} - \zeta_3 \ln \frac{t^\cut}{\mu^2}  - \frac{\pi ^4}{80}    \Bigr)
\nn\\  &\quad
+ \biggl[2C_i\delta_{ij} \mathcal{L}_0(1-x) + \hat{P}_{ij}^{(0)}(x) - \dfrac{2C_i\delta_{ij}}{1-x} \biggr]
\Bigl(  \frac{\pi^2}{12} \ln \frac{t^\cut}{\mu^2} - \frac{\pi^2}{12} \ln x  - \frac{\zeta_3}{2}  \Bigr)
\nn\\&\quad
+ C_i \, \delta_{ij} \, \frac{\pi^2}{6}   \mathcal{L}_1(1-x)
+ \frac{\pi^2}{12}\ln(1-x)\biggl[\hat{P}_{ij}^{(0)}(x) - \dfrac{2C_i\delta_{ij}}{1-x}\biggr]
\biggr\}
\,.\end{align}
The functions $\hat{P}^{(0)}_{ij}(x)$ and $\hat{P}_{ij}(x)$ are as defined in \eq{oneloopsplit}, and the $C_i$ are as defined in \eq{Ci}.

The remaining  $t^\cut$ independent corrections come from the integration of the $\mathcal{A}_B$ part of the amplitude in \eq{ampsepAB}. We decompose them as in \mycites{Gaunt:2014xga, Gaunt:2014cfa} as
\begin{align}
\Delta I_{\bar q_i \bar q_j}^\two(x, R) = \Delta I_{q_i q_j}^\two(x, R)
&= C_F\, \theta(x) \bigl[ \delta_{ij} \Delta I_{qqV}^\two(x, R) + \Delta I_{qqS}^\two(x, R) \bigr]
\,, \nn \\
\Delta I_{\bar q_i q_j}^\two(x, R) = \Delta I_{q_i \bar q_j}^\two(x, R)
&= C_F\, \theta(x) \bigl[ \delta_{ij} \Delta I_{q\bar qV}^\two(x, R) + \Delta I_{qqS}^\two(x, R) \bigr]
\,, \nn \\
\Delta I_{\bar q_i g}^\two(x, R) = \Delta I_{q_i g}^\two(x, R)
&= T_F\, \theta(x)\, \Delta I_{qg}^\two(x, R)
\,, \nn \\
\Delta I_{gg}^\two(x, R)
&= \theta(x)\, \bigl[C_A\,\Delta I_{ggA}^\two(x, R) + T_F n_f\,\Delta I_{ggF}^\two(x, R) \bigr]
\,, \nn \\
\Delta I_{g q_i}^\two(x, R) = \Delta I_{g \bar q_i}^\two(x, R)
&= C_F\, \theta(x)\, \Delta I_{gq}^\two(x, R)
\,.\end{align}
The results are
\begin{align} \label{eq:DeltaIqqV}
\Delta I_{qqV}^{\two}(x,R)
&= \delta(1-x)\, G(R)
\\& \quad
+ C_A \biggl\{
  \Bigl(\frac{1}{8} - \frac{\pi^2}{3}\Bigr) \Bigl(\hat P_{qq}(x) - \frac{2}{1-x}\Bigr) \ln\frac{R}{2\pi}
+ \frac{13}{8} - \frac{3 \ln2}{2}
\nn \\ & \qquad
+ 2\pi^2 \frac{4h_{qq}^{C_FC_A}(x)+ 11 h_{qq}^{C_FT_F}(x)}{1-x}
+ \frac{0.2^2-R^2}{4} \biggl[
     \frac{\pi^2}{6} (1 + x)
   \nn \\ & \qquad\quad
   - \frac{2611}{2400} - \frac{10919 x}{7200} + \frac{3377 x^2}{1200}
   + \Bigl(\frac{6}{5} + \frac{43 x}{30} - \frac{14 x^2}{5} \Bigr)\ln2
\biggr]
+ \mathcal{O}(R^4)
\biggr\}
\nn \\ & \quad
+ \beta_0\biggl\{
\Bigl(\frac{23}{24} - \ln2\Bigr)\Bigl(\hat P_{qq}(x)- \frac{2}{1-x}\Bigr)\ln\frac{R}{2\pi}
   - \frac{13}{24} + \frac{\ln2}{2}
-6 \pi^2 \frac{h_{qq}^{C_FT_F}(x)}{1-x}
\nn \\ & \qquad
+ \frac{0.2^2-R^2}{4} \biggl[
   -\frac{3389}{7200} + \frac{391 x}{2400} - \frac{253 x^2}{400}
   + \Bigl(\frac{13}{30} - \frac{x}{5} + \frac{3 x^2}{5}\Bigr) \ln2
\biggr] 
+ \mathcal{O}(R^4)
\biggr\}
\nn \\ & \quad
+ C_F\, 8\pi^2 h_{qq}^{C_F^2}(x)
+ (C_A - 2 C_F) \biggl\{
\frac{0.2^2-R^2}{4}
\Bigl[
   - x \hat P_{qq}(x) [L_1(x) - 2 L_2(x)]
   \nn \\ & \qquad\quad
   - 2 x^2 L_1(x)
   + \frac{115}{36} + \frac{31 x}{9} + \frac{11 x^2}{12}
   - \Bigl(\frac{10}{3} + \frac{10x}{3} + x^2\Bigr) \ln2
\Bigr]
+ \mathcal{O}(R^4)
\biggr\}
\nn
\,.\end{align}
\begin{align}
\Delta I_{q\bar{q}V}^{\two}(x, R)
&= 2 (2C_F - C_A)
\biggl\{
2 \pi^2 h_{q\bar{q}V}(x)
+ \frac{0.2^2-R^2}{4} (1-x) \Bigl[\frac{x}{1+x} L_1(x) - (1-\ln2) \Bigr]
\nn \\ & \quad
+ \mathcal{O}(R^4) \biggr\}
\,,\end{align}
\begin{align}
\Delta I_{qqS}^{\two}(x, R)
&= T_F \biggl\{
8 \pi^2 h_{qqS}(x)
+ \frac{0.2^2-R^2}{4} 
\Bigl[
   - 4 x \bigl[(1 - x)^2 L_2(x) + L_1(x) \bigr]
\nn \\ & \qquad\quad
   + \frac{95 + 49 x}{18} - \frac{8}{3} (2 + x) \ln2
\Bigr]
+ \mathcal{O}(R^4)
\biggr\}
\,,\end{align}
\begin{align}
\Delta I_{qg}^{\two}(x, R)
&= C_F \biggl\{
P_{qg}(x) \Bigl(3 - \frac{\pi^2}{3} - 3\ln{2} \Bigr) \ln\frac{R}{2\pi}
+ 8 \pi^2 h_{qg}^{C_FT_F}(x)
+ \frac{0.2^2-R^2}{4}
\biggl[
   - \frac{\pi^2}{2} P_{qg}(x)
   \nn \\ & \qquad\quad
   + \frac{23}{4} - \frac{161 x}{18} + \frac{427 x^2}{36}
   - \Bigl(\frac{11}{2} - \frac{25 x}{3} + \frac{34 x^2}{3} \Bigr) \ln2
\biggr]
 + \mathcal{O}(R^4)
\biggr\}
\nn \\ & \quad
+ C_A \biggl\{
8 \pi^2 h_{qg}^{C_AT_F}(x)
+ \frac{0.2^2-R^2}{4}
\biggl[
   \frac{\pi^2}{3} P_{qg}(x)
   - 2 x (3 - 6 x + 4 x^2) L_2(x)
   \nn \\ & \qquad\quad
   - x (1 - 2 x + 6 x^2) L_1(x)
   - \frac{41}{6} + \frac{233 x}{18} - \frac{19 x^2}{9}
   + \Bigl(7 - \frac{40 x}{3} + \frac{7 x^2}{3}\Bigr) \ln2
\biggr]
\nn \\ & \qquad
+ \mathcal{O}(R^4)
\biggr\}
\,,\end{align}
\begin{align} \label{eq:DeltaIgg}
\Delta I_{ggA}^{\two}(x,R)
&= \delta(1 - x)\,G(R)
\nn \\ & \quad
+ C_A \biggl\{
  \Bigl(\frac{1}{8} - \frac{\pi^2}{3}\Bigr) \Bigl(\hat P_{gg}(x) - \frac{2}{1-x}\Bigr) \ln\frac{R}{2\pi}
+ \frac{13}{8} - \frac{3 \ln2}{2}
\nn \\ & \qquad
+ 2 \pi^2 \frac{4h_{gg}^{C_A^2}(x) + 11 h_{gg}^{C_AT_F}(x)}{1-x}
+ \frac{0.2^2-R^2}{4} \biggl[
     x \hat P_{gg}(x) [L_1(x)-2L_2(x)]
   \nn \\ & \qquad\quad
   + 4 x (1 - x)^2 [L_2(x) - L_1(x)]
   - \frac{4 (1 -3 x - 3 x^4)}{1 + x} L_1(x)
   \nn \\ & \qquad\quad
   - \frac{\pi^2}{3}\Bigl(\frac{1}{x} -2 + x - x^2\Bigr)
   - \frac{4997}{360 x} + \frac{40933}{3600} - \frac{42521 x}{2400} - \frac{36853 x^2}{7200}
   \nn \\ & \qquad\quad
   + \Bigl(\frac{44}{3 x} - \frac{397}{30} + \frac{96 x}{5} + \frac{131 x^2}{30}\Bigr) \ln2
\biggr] + \ord{R^4}
\biggr\}
\nn \\ & \quad
+ \beta_0\biggl\{
\Bigl(\frac{23}{24} - \ln2\Bigr)\Bigl(\hat P_{gg}(x)- \frac{2}{1-x}\Bigr)\ln\frac{R}{2\pi}
   - \frac{13}{24} + \frac{\ln2}{2}
-6 \pi^2 \frac{h_{gg}^{C_A T_F}(x)}{1-x}
\nn \\ & \qquad
+ \frac{0.2^2-R^2}{4}
\biggl[
 \frac{133}{120 x} - \frac{5911}{3600} + \frac{3307 x}{2400} - \frac{2683 x^2}{2400}
   \nn \\ & \qquad\quad
   - \Bigl(\frac{1}{x} - \frac{49}{30} + \frac{7 x}{5} - \frac{11 x^2}{10} \Bigr) \ln 2
\biggr]
+ \mathcal{O}(R^4)
\biggr\}
\,,\end{align}
\begin{align}
\Delta I_{ggF}^{\two}(x,R)
&= C_F \biggl\{
8 \pi^2 h_{gg}^{C_FT_F}(x) + \frac{0.2^2-R^2}{4}\, \frac{(1-x)(2-2 x+x^2)}{9x} (23 - 24 \ln2)
\nn \\ & \quad
+ \mathcal{O}(R^4)
\biggr\}
\,,\end{align}
\begin{align}
\Delta I_{gq}^{\two}(x, R)
&= C_F \biggl\{
P_{gq}(x)\Bigl(3 - \frac{\pi^2}{3} - 3\ln{2} \Bigr)\ln\frac{R}{2\pi} + 8 \pi^2 h_{gq}^{C_F^2}(x)
+ \frac{0.2^2-R^2}{4}
\biggl[
     \frac{\pi^2}{6} P_{gq}(x)
   \nn \\ & \qquad
   - \frac{53}{3 x} + \frac{247}{18} + \frac{11x}{18} - 4 x^2
   + \Bigl(\frac{19}{x} - \frac{47}{3} + \frac{x}{6} + 4 x^2 \Bigr) \ln 2
\biggr]
 + \mathcal{O}(R^4)
\biggr\}
\nn \\ & \quad
+ C_A \biggl\{
8 \pi^2 h_{gq}^{C_FC_A}(x)
+ \frac{0.2^2-R^2}{4}
\biggl[
   - 2 (2 - 4 x + 5 x^2 - 2 x^3)\, L_2(x)
   \nn \\ & \qquad\quad
   - (2 - 10 x + x^2)\, L_1(x)
   - \frac{\pi^2}{3} P_{gq}(x)
   + \frac{29}{3 x} - \frac{199}{18} - \frac{119 x}{18} + 4 x^2
   \nn \\ & \qquad\quad
   - \Bigl(\frac{10}{x} - \frac{35}{3} - \frac{19 x}{3} + 4 x^2 \Bigr) \ln2
\biggr]
+ \mathcal{O}(R^4)
\biggr\}
\,.\end{align}
In the results above, we have defined
\begin{align}
G(R)
&= C_A \biggl[ \Bigl(\frac{1}{8} - \frac{\pi^2}{3}\Bigr) \ln^2 R + \Bigl(-\frac{49}{36} + \frac{7}{12}\ln2 + 9 \zeta_3\Bigr) \ln R
    + 4 g^{C_A}(R) + 11 g^{T_F}(R)
    \biggr]
 \nn \\ & \quad
 + \beta_0 \biggl[
 \Bigl(\frac{23}{24} - \ln2\Bigr) \ln^2 R + \Bigl(-\frac{17}{3} + \frac{\pi^2}{3} + \frac{23}{12}\ln2 + \ln^2 2\Bigr) \ln R
 - 3 g^{T_F}(R)
 \biggr]
\,,\end{align}
and the functions $L_1(x)$ and $L_2(x)$ are defined as
\begin{align}
L_1(x)
&= \frac{1}{1-x}\Bigl[\frac{\pi^2}{6} - \mathrm{Li}_2(x^2) - 2 \ln(1-x) \ln x \Bigr]
= 2(1 - \ln2) + (1 - \ln2)(1 - x) + \dotsb
\,,\nn\\
L_2(x) &= \frac{L_1(x) + 2\ln(1+x)}{1-x} + 2\frac{x \ln x}{(1-x)^2}
= 1 - \ln2 + \Bigl(\frac{13}{18} - \frac{2}{3}\ln2\Bigr) (1-x) + \dotsb
\,,\end{align}
and as shown are regular for $x\to 1$.

The functions $g(R)$ and $h(x)$ are fitted using numerically generated data points as explained in the beginning of this section.
The functions $g(R)$ give the nonlogarithmic $R$-dependence of the $\delta(1-x)$ terms in $\Delta I_{gg}$ and $\Delta I_{qqV}$.
They are fitted using the form in \eq{fitform_f}, and the fitted coefficients are given in table \ref{tab:g_BF_coeff}.

\begin{table}[t]
\centering
\begin{tabular}{c|c c c c c}
\hline\hline
Channel & $g_1$ & $g_2$ & $g_3$ & $g_4$ & $g_5$\\ \hline
$C_A$ & 0.12595  & $-0.13881$ & $-2.0197\cdot 10^{-2}$  & 0.16527 & $2.5289\cdot 10^{-3}$
\\
$T_F$ & $-0.27276$ & $8.3507\cdot10^{-3}$ & $-9.4394\cdot10^{-4}$ & $5.0721\cdot10^{-3}$ & $6.7609\cdot10^{-4}$
\\\hline\hline
\end{tabular}
\caption{
Fit function coefficients $g_i$ for the correlated soft function $g(R) = g_1 + g_2 R^2 + g_3 R^4 + g_4 R^2\ln R + g_5 R^4\ln R $.
\label{tab:g_BF_coeff}}
\end{table}
 
The function $h(x)$ gives the $x$-dependence of the $\epsilon^0$ non-$\delta(1-x)$ piece in $\Delta I$ at the point $R=0.2$. 
Where appropriate we subtracted out the analytically-calculated small-$R$ result and/or the endpoint $x=1$ behaviour. 
For this function, a more complicated form is used. In fact, it is fitted separately in two regions: $0.0009 < x < 0.6$ and $0.6 < x < 1$, where $x=0.0009$ is the smallest $x$ value for which we generate data. In the high-$x$ region $0.6 < x < 1$, we use a form that is equal to the first few terms of a Taylor expansion around $x=1$.
In the low-$x$ region $0.0009 < x < 0.6$, we use a form equal to the first few terms of a Taylor expansion around $x=0$, plus terms with powers of $\ln x$ up to the third power.
For the $gg$ and $gq$ channels only, a final term proportional to $1/x$ is also included. It can be shown that only these channels have a piece in $h(x)$ proportional to $1/x$ by expanding the respective integrands for small $x$.
Summarizing, we use the following form to fit $h(x)$:
\begin{align} \label{eq:fitform_h}
 h(x) = \begin{cases}
         \sum_{n=-1,6} d_n x^n + d_7\ln x + d_8\ln^2 x + d_9\ln^3x \quad & 0.0009 < x < 0.6 \,,
         \\
         \sum_{n=0,7} \tilde{d}_n (1-x)^n  \quad & 0.6 < x < 1 \,.
        \end{cases}
\end{align}
This functional form is satisfactory for every $h(x)$ function in \eqst{DeltaIgg}{DeltaIqqV}. The only exception is $h_{qqS}$, which falls to zero so steeply near $x=1$ that it cannot be described well by the above high-$x$ fit function. For this function only we therefore use a high-$x$ fit function given by $\sum_{n=3,9} \tilde{d}_n (1-x)^n$.
Including these higher powers of $(1-x)$ ameliorated the fit, and in this case including the lower powers of $(1-x)$ up to the second power is not necessary for a good fit.

We have performed the fits for the coefficients $d_n$ and $\tilde{d}_n$  separately for each $h(x)$, each using $30$ data points in the relevant $x$ range. 
We checked that the resulting fit reproduces all the data points at the sub-per-mille level, and confirmed by eye that it does not contain interpolation artifacts.
We also checked by eye that the low-$x$ fit transitions smoothly onto the high-$x$ fit.
The fit coefficients for all $h(x)$ functions are presented in \app{BFfitcoff}.

As mentioned in section \ref{subsec:beam} and indicated above, our results for $\Delta I_{ij}$ only contain terms up to $\mathcal{O}(R^2)$ for the non-$\delta(1-x)$ pieces.
For the reader interested in the more precise behaviour in $R$ of these pieces, we have also obtained full 2D grids by numerically integrating the beam function amplitude minus its small $\Delta R$ expansion, spanning $0.0009 < x < 1$ and $0.2 < R < 1.2$.
These results can be provided upon request.

For most parton and color channels, the level of error of the $\mathcal{O}(R^2)$-accurate expressions above compared to the full numerical results is a few permille for $R$ values up to around $0.6-0.8$, and a few percent at the largest $R$ values $\sim 1.0-1.2$.
The exceptions to this are the $qqV$ $C_F^2$ and $gg$ $C_A^2$ channels, where the errors hit the percent level already for $R=0.6$, and are much bigger for larger $R$ values.%
\footnote{We also see a similar situation for the $q\bar{q}V$ piece at large $x$ values, but since its overall contribution to the beam function is so tiny that the error should be irrelevant.}
For these two channels we have also computed the $R^4$ corrections, and their inclusion improves the agreement with the full numerical result to a similar level as the other channels.
The expressions for these $R^4$ terms may also be provided upon request.

\subsection{Independent emission pieces} \label{subsec:resultsindepem}

The corrections to the two-loop soft and beam functions associated with the clustering of independent emissions, together with the soft-collinear mixing term, are given by
\begin{align}
\label{eq:indepS}
\Delta{S}_{f,\mathrm{indep}}^{(2)}(\Tau^{\cut}, R, \mu)
&= 16C_i^2\, R^2  \biggl[\frac{\pi^2}{3}\ln\frac{\Tau^\cut}{\mu} - 2 \zeta_3
+ \frac{\pi^2}{6}\mathbb{F}(R) +  \mathcal{U}_f(R) \biggr]
\,,\\
\label{eq:indepI}
\Delta{\cI}^\two_{ij,\mathrm{indep}}(t^{\cut}, x, R, \mu)
&= 4 C_i \, \hat{P}^{(0)}_{ij}(x)(1-x)
\\\nn &  \quad \times
R^2  \biggl[ \frac{\pi^2 }{6} \mathcal{L}_0(1-x) + \delta(1-x) \biggl(\frac{\pi^2 }{6} \ln \frac{t^\cut}{\mu^2}
+\frac{\zeta_3}{4} + \frac{\pi^2}{6}\mathbb{F}(R) \biggr)\biggr]
\,, \\
\label{eq:SCmixing}
S\hspace{-0.5mm}C^\two_{ij}(t^{\cut},z,R, \mu)
&= -8C_i\,\hat{P}^{(0)}_{ij}(x) (1-x)
\\\nn &  \quad \times
R^2 \biggl[ \frac{\pi^2}{6} \mathcal{L}_0(1-x) + \delta(1-x) \biggl(\frac{\pi^2}{6} \ln \frac{t^\cut}{\mu^2}
- \zeta_3 + \frac{\pi^2}{6}\mathbb{F}(R) \biggr)\biggr]
\,,\end{align}
where
\begin{align} \label{eq:indepseries}
\mathbb{F}(R)
&= -\frac{1}{2} + \ln R - \frac{1}{6}\Bigl(\frac{R}{2}\Bigr)^2 - \frac{1}{90} \Bigl(\frac{R}{2}\Bigr)^4  - \frac{1}{567}\Bigl(\frac{R}{2}\Bigr)^6 + \mathcal{O}(R^8)
\\
\mathcal{U}_B(R)
&= - \Bigl(\frac{R}{2}\Bigr)^2 - \frac{64}{45 \pi} \Bigl(\frac{R}{2}\Bigr)^3 - \frac{1}{9} \Bigl(\frac{R}{2}\Bigr)^4 + \frac{1}{135}\Bigl(\frac{R}{2}\Bigr)^6
- \frac{1}{945}\Bigl(\frac{R}{2}\Bigr)^8 + \ord{R^{10}}
\,, \\
\mathcal{U}_C(R)
&= -2 \Bigl(\frac{R}{2}\Bigr)^2 - \frac{2}{9}\Bigl(\frac{R}{2}\Bigr)^4 + \frac{2}{135}\Bigl(\frac{R}{2}\Bigr)^6
- \frac{2}{945}\Bigl(\frac{R}{2}\Bigr)^8 + \ord{R^{10}}
\,.\end{align}
$\mathcal{U}_B(R)$ and $\mathcal{U}_C(R)$ are the terms associated with the measurements $\Delta\cM_{B,\indep,2}$ and $\Delta \cM_{C,\indep,2}$ in \eqs{DeltaMBuncorr2}{DeltaMCuncorr2}, respectively, cf. \app{ressoftuncorr}. Further terms in the $R$ expansion may be easily computed, but have a negligible impact for $R<1.5$. From the above expansions it seems clear that the expansion parameter is $R/2$ or even smaller.

As mentioned already, the factorization of the jet-veto measurement, does not hold at $\ord{R^2}$ due to the soft-collinear mixing contributions. As a result, the scale dependence from the independent emission terms only cancels, when all three contributions in \eqss{indepS}{indepI}{SCmixing} are correctly combined in the cross section, see \eq{sigmaRsub}. (Equivalently, the $1/\epsilon$ divergences in the corresponding bare contributions only cancel between all three contributions.) They therefore have to be included together in a consistent way.

There are two possibilities as to how one can treat these $\ord{R^2}$ terms at NNLL$'$.
These different treatments are formally equivalent at NNLL but not necessarily beyond.
In the first, we note that at the two-loop order we work here, the soft-collinear term has the same type of logarithms as the beam function. This allows us, at this order at least, to absorb the soft-collinear mixing terms into the beam functions, such that in \eqs{masterS}{masterI} we have
\begin{align} \label{eq:uscheme1}
\Delta\mathbb{I}_{ij}^{\two}(t^\cut,x,\mu)
&= \Delta{I}^\two_{ij,\mathrm{indep}}(t^{\cut},x,R,\mu) + S\hspace{-0.5mm}C^\two_{ij}(t^{\cut},x,R,\mu)
\,,\nn \\
\Delta\mathbb{S}_f^\two(\Tau^{\cut},R,\mu)
&= \Delta S_{f, \indep}^{\two}(\Tau^{\cut},R,\mu)
\,,\end{align}
and the $\mu$-dependence cancels between $\Delta \mathbb{I}_{ij}^\two$ and $\Delta \mathbb{S}_f^\two$.
In practice, this scheme is effectively equivalent to the one employed in \mycite{Becher:2013xia} for a $p_{Tj}$ veto.

Let us define the anomalous dimensions of the beam and soft functions (for either $\Tau_{Bj}$ or $\Tau_{Cj}$) as
\begin{align}
\gamma_S(\Tau^\cut, \mu, R)
&= \gamma_{G,S}(\Tau^\cut, \mu) + \Delta \gamma_S[\alpha_s(\mu), R]
\,,\nn \\
\gamma_B(t^\cut, \mu)
&= \gamma_{G,B}(t^\cut, \mu) + \Delta \gamma_B[\alpha_s(\mu), R]
\,,\end{align}
where $\gamma_{G,S}$ is the cumulant of the anomalous dimension for thrust (and C-parameter), and $\gamma_{G,B}$ is the cumulant of the anomalous dimension for the virtuality-dependent beam function. Then in the scheme defined by \eq{uscheme1}, the jet clustering corrections to the (noncusp) anomalous dimensions are given by
\begin{align} \label{eq:deltaGamma}
\Delta \gamma^i_{S}(\alpha_s, R)
= -2\Delta \gamma^i_{B}(\alpha_s, R)
=  \Bigl(\frac{\alpha_s}{4\pi}\Bigr)^2 \Bigl[ \Delta\gamma_{S\,1}^i(R) - C_i^2 \frac{16\pi^2}{3} R^2
\Bigr]
\,,\end{align}
where the last $C_i^2 R^2$ term is from the independent emission contributions.

Alternatively, we can simply exclude \eqss{indepS}{indepI}{SCmixing} from the factorized cross section, in which case we have in \eqs{masterS}{masterI}
\begin{equation} \label{eq:uscheme2}
\Delta\mathbb{I}_{ij}^{\two}(t^\cut,x) =0
\,,\qquad
\Delta\mathbb{S}_{f}^{(2)} (\Tau^{\cut}) = 0
\,,\end{equation}
and in addition the last $C_i^2 R^2$ term is removed from \eq{deltaGamma}.
Instead, we combine all terms into a common independent emissions contribution and add the following correction term to the $0$-jet cross section:
\begin{equation} \label{eq:sigmaRsub}
\Delta \sigma_0^{\mathrm{Rsub}}
= \frac{\alpha_s^2}{(4\pi)^2}H^{(0)}\biggl[\delta(1-x)\delta_{ij} \Delta S_{f,\indep}^{\two} + 2\Delta \cI_{ij,\indep}^{\two}(x) + 2S\hspace{-0.5mm}C^\two_{ij}(x) \biggr] \otimes f_i(x)\, f_j(x)
\,.\end{equation}
This equation is somewhat schematic -- the $\otimes$ symbols represent the Mellin convolution in light-cone momentum fractions along with the appropriate flavour sums, and all pieces are evaluated at a common scale $\mu$. In the resummed cross section this term can then be multiplied with the overall evolution factor of the $\ord{R^0}$ cross section. This corresponds to what is done in \mycites{Banfi:2012jm, Stewart:2013faa} in the context of a $p_{Tj}$ veto.

\section{Conclusion}
\label{sec:conclusion}

In this paper, we have computed the two-loop beam and dijet soft functions for the rapidity-dependent jet-veto observables $\Tau_{Bj}$ and $\Tau_{Cj}$. Each function is computed as the difference from the known results for the corresponding global jet-independent observable. The beam and soft functions have been computed for the complete set of color and parton channels. The dominant parts of the functions in the small-$R$ limit are computed fully analytically -- e.g. the $\ln^n R$ terms and the $R^2$ terms in the non-$\delta(1-x)$ part of the beam functions. The remaining parts are determined in terms of finite numerical integrals, for which we provide simple interpolation functions.

Our results enable the resummation of color-singlet production cross sections with $\Tau_{Bj}$ or $\Tau_{Cj}$ jet vetoes at full NNLL$'$ order, and are a necessary ingredient for the N$^3$LL resummation (with the missing ingredient being the corrections to the three-loop anomalous dimension). Our computed beam functions can also be used in the resummation of $N$-jet cross sections for which a central jet veto is imposed using $\Tau_{Bj}$ or $\Tau_{Cj}$.
Such measurements will provide important information on the production pattern of additional jets in hard processes at the LHC.

\begin{acknowledgments}
This work was supported by the DFG Emmy-Noether Grant No. TA 867/1-1. J.G.
acknowledges financial support  from  the  European  Community  under  the  FP7  Ideas  program  QWORK (contract  320389). 
M.S. is supported by GFK and by the PRISMA cluster of excellence at Mainz university.
The authors thank the Erwin Schr{\"o}dinger Institute program ``Challenges and Concepts for Field Theory and Applications in the Era of the LHC'', and J.G. thanks JGU Mainz, for hospitality while portions
of this work were completed. The figure in this paper was produced using JaxoDraw \cite{Binosi:2003yf}. 
\end{acknowledgments}

\appendix

\section{Soft function correction: independent emission clustering part}
 \label{sec:ressoftuncorr}

\subsection[\texorpdfstring{$\Tau_{Bj}$}{TauBj} veto measurement]
{$\Tau_{Bj}$ veto measurement}
\label{sec:TauBuncorr}

We first compute the contribution to the soft function corresponding to $\Delta \mathcal{M}_{B,\mathrm{indep,1}}$ in \eq{DeltaMBuncorr1}.
We have
 \begin{align} \label{eq:deltaSBuncorr}
\Delta S^\two_{B,\mathrm{indep},1}
&=  g^4\Bigl(\frac{\mu^2 e^{\gamma_E}}{4\pi}\Bigr)^{2\epsilon}\int\!\frac{\df^d k_1}{(2\pi)^d}
\,\frac{\df^d k_2}{(2\pi)^d} \, (2\pi)\delta^+(k_1^2)
(2\pi)\delta^+(k_2^2) \, \mathcal{A}_{\mathrm{indep}}(k_1,k_2) \, \theta(\Delta R < R)
 \nn \\  &\quad
\times 2 \theta(y_t>0)  \bigl[ \theta(k_1^+ + k_2^+ < \Tau^\cut) - \theta(k_1^+ < \Tau^\cut) \theta(k_2^+ < \Tau^\cut) \bigr]
\end{align}
where
\begin{align}
 \mathcal{A}_{\mathrm{indep}}(k_1,k_2) & = \frac{8C_F^2}{k_1^+k_2^+k_1^-k_2^-}\,.
\end{align}
We express this integral in terms of light-cone momentum components, perform the transverse integrals using the delta functions, and change variables from $k_1^+,k_2^+,k_1^-,k_2^-$ to $k_1^+,k_2^+,y_t,\Delta y$, with $y_t$ and $\Delta y$ defined in \eq{variablesTauB}. This gives
\begin{align} \label{eq:DeltaSB1uncorr2}
\Delta S^\two_{B,\mathrm{indep},1}
 &= \frac{16g^4}{(2\pi)^{2(d-1)}}  \Bigl(\frac{\mu^2 e^{\gamma_E}}{4\pi}\Bigr)^{2\epsilon}
\frac{\pi^{1-\epsilon}}{\Gamma[1-\epsilon]}
\frac{\pi^{\tfrac{1}{2}-\epsilon}}{\Gamma[\tfrac{1}{2}-\epsilon]} C_F^2
\nn \\ & \quad\times
\int\!   \df \Delta y
\,\df k_1^+ \,\df k_2^+ \,\df \Delta \phi  \,\df y_t \,\theta(y_t>0)\, e^{-4y_t\epsilon} \; (k_1^+ k_2^+)^{-1-\epsilon}  \sin^{-2\epsilon}( \Delta \phi ) 
\theta(\Delta R   < R)
\nn\\ & \qquad\times
\bigl[ \theta(k_1^+ + k_2^+ < \Tau^\cut) - \theta(k_1^+ < \Tau^\cut) \theta(k_2^+ < \Tau^\cut) \bigr]
\nn \\
&= \Bigl( \frac{\alpha_s}{\pi} \Bigr)^2  C_F^2 R^2 \Bigl( \frac{\mu}{\Tau^\cut}\Bigr)^{4\epsilon}
\frac{\pi^2}{6}\biggl[-\frac{1}{2 \epsilon }-\frac{12 \zeta_3}{\pi ^2} + \mathbb{F}(R) \biggr]
\,.\end{align}

For the calculation of $\Delta S_{B,\mathrm{indep},2}$ we use the  variables in \eq{variablesTauB} and express the measurement as
\begin{align}
\Delta \mathcal{M}_{B,\mathrm{indep,2}}
&=2 \theta(\Delta R   < R) \,\theta\Bigl(0 < y_t< \Big| \frac{\Delta y}{2} \Big| \Bigr)
\,\theta\biggl[ \Tau^\cut <\Tau_T < \frac{\Tau^\cut}{\max(z,1-z)} \biggr]
\nn\\ &\quad
-4 \theta(\Delta R   < R) \, \theta(\Delta y>0)  \, \theta\Bigl( 2y_t -\Delta y + \ln\Bigl[ 1+ \bigl( e^{2\Delta y} -1  \bigr) z  \Bigr]  \Bigr)
\nn\\ &\qquad \times
\theta\Bigl(- \frac{\Delta y}{2} < y_t< \frac{\Delta y}{2} \Bigr) \,\theta\biggl[ \Tau^\cut <\Tau_T < \Tau^\cut  \min \Bigl( \frac{1}{z}, \frac{e^{\Delta y - 2 y_t}}{1-z} \Bigr) \biggr]
\,.
\end{align}
After performing the $\Tau_T$, $\Delta \phi$, $y_t$  integrations we expand in small $\Delta y \sim R$ and finally integrate over $z$ and $\Delta y$.
The result is given in \subsec{resultsindepem}.

\subsection[\texorpdfstring{$\Tau_{Cj}$}{TauCj} veto measurement]
{$\Tau_{Cj}$ veto measurement}
\label{sec:TauCuncorr}

For $\Tau_C$, the starting expression is identical to \eq{deltaSBuncorr}, albeit without the $2\theta(y_t>0)$ piece, and with $k_1^+$ and $k_2^+$ replaced by $\Tau_{C1}$ and $\Tau_{C2}$ in the theta functions of the second line. We change variables in this case from $k_1^+,k_2^+,k_1^-,k_2^-$ to $\Tau_{C1},\Tau_{C2}, y_t, \Delta y$, with
\begin{align}
 \Tau_{Ci} = \frac{k_i^+ k_i^-}{k_i^+ + k_i^-} \,, 
\end{align}
yielding
\begin{align}
\Delta S^\two_{C,\mathrm{indep},1}
&= \frac{8g^4}{(2\pi)^{2d-2}}  \Bigl(\frac{\mu^2 e^{\gamma_E}}{4\pi}\Bigr)^{2\epsilon}
\frac{\pi^{1-\epsilon}}{\Gamma[1-\epsilon]}
\frac{\pi^{\tfrac{1}{2}-\epsilon}}{\Gamma[\tfrac{1}{2}-\epsilon]}\, C_F^2 \int_{-\infty}^\infty \! \df y_t \bigl[ 2+2\cosh(2y_t) \bigr]^{-2\epsilon}
\nn \\ &\quad \times
\int \! \df \Delta y \, \df \Tau_{C1} \,\df \Tau_{C2} \, \df \Delta \phi  \,
(\Tau_{C1} \Tau_{C2})^{-1-\epsilon}  \sin^{-2\epsilon} ( \Delta \phi) \, \theta(\Delta R   < R)
\nn \\ &\qquad \times
\Bigl[ \theta(\Tau_{C1} + \Tau_{C2} < \Tau^\cut) - \theta(\Tau_{C1} < \Tau^\cut) \theta(\Tau_{C1} < \Tau^\cut) \Bigr]
\nn \\
&= \Delta S_{B,\mathrm{indep},1} + \ord{\e}\,.
\end{align}

The measurement function for $\Delta S_{C,\mathrm{indep},2}$ in \eq{DeltaMCuncorr2} can be written as ($0<X<1$)
\begin{align}
 \Delta \mathcal{M}_{C,\mathrm{indep,2}} &=  - \theta(\Delta R < R) \, \theta \Bigl( X \Tau^\cut   <   \Tau_T^\prime  <  \Tau^\cut \Bigr)\,,
\end{align}
with
\begin{align}
 X = \frac{\Tau_T^\prime}{\Tau_{Cj}} =
2 e^{2 ( \Delta y+y_t)} \frac{ (2 z^\prime-1) \sinh ( \Delta y) \sinh (2 y_t)+\cosh ( \Delta y) \cosh (2 y_t)+1 }{\big( e^{ \Delta y+2 y_t} + e^{2  \Delta y} (1-z^\prime)+z^\prime \big)
\big(e^{2 y_t} \big[ (e^{2  \Delta y}-1) z^\prime+1 \big]+e^{ \Delta y}\big)} \,,
\end{align}
and
\begin{align}
\Tau_T^\prime = \Tau_{C1}+ \Tau_{C2}\,, \qquad z^\prime = \frac{\Tau_{C1}}{\Tau_T^\prime}\,.
\label{eq:variablesTauC}
\end{align}
We can now directly carry out the intergrations over $\Tau_T^\prime$ and $\Delta \phi$. After expanding in small $\Delta y \sim R$ we then integrate over $y_t$, $z^\prime$, and $\Delta y$ to arrive at the result given in \subsec{resultsindepem}.

\section{\boldmath Soft function correction: \texorpdfstring{$\Delta S_{\rm rest}$}{Delta Srest}}
\label{sec:DeltaSrestCalc}

For completeness we document in this appendix how we organize the numerical computation of $\Delta S^\two_{{\rm rest}, f}$ in \eq{StotBC}.
In particular we show the split of the corresponding measurement function $\Delta \mathcal{M}_{{\rm rest}, f}$, we have chosen in order to conveniently divide the calculation into separate parts.
We note however that other decompositions (or no split at all) might work just as well in practice.

For the $\Tau_{Bj}$ veto we write
\begin{align}
  \Delta \mathcal{M}_{{\rm rest}, B} & = \Delta \mathcal{M}_{{\rm rest}, B1} + \Delta \mathcal{M}_{{\rm rest}, B2} + \Delta \mathcal{M}_{{\rm rest}, B3}\,,
\end{align}
where
\begin{align}
\Delta \mathcal{M}_{{\rm rest}, B1} &= 4\theta(\Delta R < R)\,\theta(y_j>0)\,\theta(y_1>0)\,\theta(y_2<0)
\nn\\ &\quad \times
\Bigl[ \theta(k_1^+ + k_2^+ < \Tau^\cut) - \theta(k_1^+ < \Tau^\cut)\, \theta(k_2^- < \Tau^\cut) \Bigr]
\nn\,,  \\
\Delta \mathcal{M}_{{\rm rest}, B2} &= 2\theta(\Delta R > R) \Bigl[ \theta(y_1>0)\,\theta(y_2>0) - \theta(y_t>0) \Bigr]
\nn\\ &\quad \times
\Bigl[ \theta(k_1^+ < \Tau^\cut)\, \theta(k_2^+ < \Tau^\cut)  -\theta(k_1^+ + k_2^+ < \Tau^\cut) \Bigr]
\,,  \\\nn
\Delta \mathcal{M}_{{\rm rest}, B3}
&= 2\theta(y_1>0)\,\theta(y_2<0) \Bigl[ \theta(k_1^+ < \Tau^\cut)\, \theta(k_2^- < \Tau^\cut) - \theta(k_1^+ + k_2^- < \Tau^\cut) \Bigr]
\,.\end{align}
The numbers (4 or 2) in front of the $\theta$ functions indicate that we have exploited phase-space symmetries and summed over equivalent contributions from different regions.
We note that $\Delta \mathcal{M}_{{\rm rest}, B3}$ amounts to the difference between the (inclusive) cumulant beam thrust and cumulant double-hemisphere soft function measurements.
Correspondingly, $\Delta S^\two_{{\rm rest}, B3}$ is a finite $R$-independent constant. A simple dimensional analysis also shows that it is independent of $\Tau^\cut$.
The piece $\Delta S^\two_{{\rm rest}, B1}$ is of $\ord{R}$ and  $\Delta S^\two_{{\rm rest}, B2}$ contains $\ord{R^0}$ terms.

For the $\Tau_{Cj}$ veto we write
\begin{align}
  \Delta \mathcal{M}_{{\rm rest}, C} & = \Delta \mathcal{M}_{{\rm rest}, C1} + \Delta \mathcal{M}_{{\rm rest}, C2}\,,
\end{align}
where
\begin{align}
\Delta \mathcal{M}_{{\rm rest}, C1} &= \theta(\Delta R > R)\, \Bigl[ \theta(\Tau_{C1} < \Tau^\cut) \, \theta(\Tau_{C2} < \Tau^\cut) -  \theta(\Tau_{C1} + \Tau_{C2} < \Tau^\cut)  \Bigr] - \Delta \mathcal{M}_{\rm base}
\nn \\
&= \theta(\Delta R > R)\,2\theta(y_t>0) 
\biggl\{
\theta \biggl[ \Tau^\cut < \Tau^\prime_T <   \frac{\Tau^\cut}{\max(z^\prime,1-z^\prime)}  \biggr]
\nn\\ &\qquad
-\theta \biggl[  \frac{\Tau^\cut}{A+B} < \Tau^\prime_T < \frac{\Tau^\cut}{\max(A,B)} \biggr]
\biggr\} \,,
\label{eq:DeltaMrestC1}
\\
\Delta \mathcal{M}_{{\rm rest}, C2} &=  \Delta \mathcal{M}_{C,\mathrm{indep,2}}  \,.   
\end{align}
Here we have defined $A = k_1^+/\Tau^\prime_T = z^\prime (1 + e^{-\Delta y - 2 y_t})$ and $B = k_2^+/\Tau^\prime_T = (1-z^\prime)(1+ e^{\Delta y-2 y_t}) $.
The corresponding part $\Delta S^\two_{{\rm rest}, C2}$  turns out to be of $\ord{R^2}$, whereas $\Delta S^\two_{{\rm rest}, C1}$ is of $\ord{R^0}$.

The form of the different parts of the measurements in $\Delta \mathcal{M}_{{\rm rest},f}$ already explains why $\Delta S^\two_{{\rm rest},f}$ does not contain terms proportional to $\ln^n R$.
For the measurements $\Delta \mathcal{M}_{{\rm rest}, B1}$ and $\Delta \mathcal{M}_{{\rm rest}, C2}$ this is obvious, because they are proportional to $\theta(\Delta R < R)$.
A term in the integrand proportional to $1/\Delta R^2$, which could potentially generate such a logarithm, must be absent, because otherwise it would cause a divergence for $\Delta R \to 0$.
The measurements $\Delta \mathcal{M}_{{\rm rest}, B2}$ and $\Delta \mathcal{M}_{{\rm rest}, C1}$ on the other hand effectively vanish linearly in the small $\Delta y \sim \Delta R$ limit. 
To see this for the latter measurement one has to take into account that in four dimensions the associated integrand is proportional to $1 / \Tau_T^\prime$. We are therefore free to rescale $\Tau^\prime_T \to \Tau^\prime_T/(A+B)$ in the second term of \eq{DeltaMrestC1}.
As the two-loop soft function integrands contain at most $1/\Delta R^2$ poles, these measurements thus prohibit logarithmic terms in $R$.

The integrations for $\Delta S^\two_{{\rm rest}, f}$ are carried out in the respective variables, see \eqs{variablesTauB}{variablesTauC}.
In fact we performed the $\Tau_T^{(\prime)}$ and $\Delta \phi$ integrations analytically and the remaining three-dimensional integral numerically.

\section{Beam function correction: independent emission clustering part} 
\label{sec:zerobins}

We first calculate the $\ord{R^2}$ contributions associated with the $\theta(\Delta R   < R) \times \mathcal{A}_A$ term in \eq{thetaRsplit}. We have
\begin{align}
& -g^4\left(\frac{\mu^2 e^{\gamma_E}}{4\pi}\right)^{2\epsilon}\int   \frac{\df^d k_1}{(2\pi)^d}
\frac{\df^d k_2}{(2\pi)^d} \,
(2\pi)\delta^+ (k_1^2) \, (2\pi)\delta^+ (k_2^2) \,
\mathcal{A}_A(k_1,k_2)
\\ \nonumber
&\qquad  \times \delta [k_1^- + k_2^- = (1-x)p^-] \, \theta \bigg[ \mathcal{T}^{\cut} < \mathcal{T}_T < \frac{\mathcal{T}^{\cut}}{\max(z,1-z)} \bigg] \,
\theta(\Delta R   < R) 
\label{eq:UncorrRsq1}
\\
&\quad = -\frac{g^4}{(2\pi)^{2(d-1)}} \left(\frac{\mu^2 e^{\gamma_E}}{4\pi}\right)^{2\epsilon}
\frac{\pi^{1-\epsilon}}{\Gamma[1-\epsilon]}
\frac{\pi^{\tfrac{1}{2}-\epsilon}}{\Gamma[\tfrac{1}{2}-\epsilon]} C_i \, \hat{P}^{(0)}_{ij}(x)  \, (1-x) p^-
\nn \\
&\qquad \times \int  
\df k_1^+ \df k_1^- \df k_2^+ \df k_2^-  \df \Delta \phi \;  (k_1^+ k_2^+ k_1^-
k_2^-)^{-1-\epsilon} \sin^{-2\epsilon}( \Delta \phi  ) \;
\theta(\Delta R   < R)
\nn \\
&\qquad \times \delta [k_1^- + k_2^- = (1-x)p^-] \, \theta \bigg[
\mathcal{T}^{\cut} < \mathcal{T}_T < \frac{\mathcal{T}^{\cut}}{\max(z,1-z)}
\bigg]
\\
&\quad = -\frac{g^4}{(2\pi)^{2(d-1)}} \left(\frac{ e^{\gamma_E}}{4\pi}\right)^{2\epsilon}  \frac{\pi^{1-\epsilon}}{\Gamma[1-\epsilon]}
\frac{\pi^{\tfrac{1}{2}-\epsilon}}{\Gamma[\tfrac{1}{2}-\epsilon]} C_i \hat{P}^{(0)}_{ij}(x) \,(1-x)\,  \bigg(\frac{\mu^2}{\Tau^{\cut}p^-}\bigg)^{2\epsilon}
\nn \\  
&\qquad \times \int  
 \df \Delta y \,  \df \Delta \phi \,   \df z \, \sin^{-2\epsilon} ( \Delta \phi )  \, \theta(\Delta R   < R) \,  2 e^{-2 \Delta y   \epsilon } \, \big[ (e^{2 \Delta y  }-1 ) z+1\big]^{2 \e} 
\nn \\
&\qquad  \quad \times  
\big[(1-x) (1-z) z \big]^{-1-2 \epsilon } 
\;
 \frac{1- \min(z,1-z)^{2\epsilon}}{2\epsilon}
\,.
\label{eq:UncorrRsq2}
\end{align}
Let us first consider the evaluation of the $1/\epsilon$ piece of 
this expression. For this purpose many of the terms in \eq{UncorrRsq2} can be set to $1$, and we obtain
\begin{align} \label{eq:naiveRsq}
\frac{1}{\epsilon} \left(\frac{\mu^2}{t^{\cut}}\right)^{2\epsilon} \lim_{x \to 1} [\hat{P}^{(0)}_{ij}(x)\,(1-x)] C_i \,  \delta(1-x)
\frac{\pi^2}{48} \Bigl( \frac{\alpha_s}{\pi} \Bigr)^2  R^2 \,,
\end{align}
where again $t^\cut \equiv xp^-\Tau^\cut$.
Plugging in the values for $C_i$ and $\hat{P}^{(0)}_{ij}(x)$ from \eqs{Ci}{oneloopsplit}, one observes that the divergent
part is zero when $i \neq j$, and for the case of $i = j$ corresponds to half 
of eq. (50) in \mycite{Tackmann:2012bt}, but with the opposite sign, as stated in the main text.

Now let us move to consider the zero-bin subtractions~\cite{Manohar:2006nz}. There are several zero
bins one must consider: $k_1$ alone soft, $k_2$ alone soft, and both
$k_1$ and $k_2$ soft simultaneously. The first two we must subtract, and the
final one we add. Our procedure for computing the zero bins coincides with
that of \mycites{Tackmann:2012bt, Jouttenus:2009ns}. 
Namely, we take the appropriate soft limit of the full amplitude $\mathcal{A}$, and also the appropriate soft limit of the rest of the integrand. This limit affects the minus momentum delta function but leaves the measurement function \eq{clusteringMbeam} unchanged.
As mentioned in the main text, taking the soft limit of 
the amplitude $\mathcal{A}$ in fact picks out only the term $\mathcal{A}_A$ 
for any of the zero bins.

We now make the decomposition in \eq{thetaRsplit}. However, for any of
the zero bins, the `$1$-term' contains scaleless integrals and is therefore zero.
For example, the `$1$-term' for the $k_1$ soft zero bin is proportional to
\begin{align} \label{eq:1termminusZB}
 \int \! dk_1^- dk_2^- \, (k_1^- k_2^-)^{-1-\epsilon} \, \delta [ k_2^- = (1-x)p^-] = 0 \,.
\end{align}
Next we consider the $\theta(\Delta R   < R)$ term. 
For the $k_1$ soft zero bin it equals
\begin{align}
& -\frac{g^4}{(2\pi)^{2(d-1)}} \left(\frac{\mu^2 e^{\gamma_E}}{4\pi}\right)^{2\epsilon}
\frac{\pi^{1-\epsilon}}{\Gamma[1-\epsilon]}
\frac{\pi^{\tfrac{1}{2}-\epsilon}}{\Gamma[\tfrac{1}{2}-\epsilon]} \, C_i \hat{P}^{(0)}_{ij}(x) \, (1-x) p^-
\nn \\
&\quad \times \int \!
\df k_1^+ \df k_1^- \df k_2^+ \df k_2^- \df \Delta \phi 
\;  (k_1^+ k_2^+ k_1^-k_2^-)^{-1-\epsilon} \; \sin^{-2\epsilon}( \Delta \phi ) \;\theta(\Delta R   < R)
\nonumber \\
&\qquad \times \delta [k_2^- = (1-x)p^-]  \; \theta  \bigg[ \mathcal{T}^{\cut} <
\mathcal{T}_T < \frac{\mathcal{T}^{\cut}}{\max(z,1-z)} \bigg]
\nn\\
&\quad = -\frac{g^4}{(2\pi)^{2(d-1)}}  \left(\frac{e^{\gamma_E}}{4\pi}\right)^{2\epsilon} \frac{\pi^{1-\epsilon}}{\Gamma[1-\epsilon]}
\frac{\pi^{\tfrac{1}{2}-\epsilon}}{\Gamma[\tfrac{1}{2}-\epsilon]} \, C_i \hat{P}^{(0)}_{ij}(x) \,(1-x) \, \bigg(\frac{\mu^2}{\Tau^{\cut}p^-}\bigg)^{2\epsilon}
\\ 
&\qquad \times \int  \!
 \df \Delta y \,  \df \Delta \phi \,  \df z \; \sin^{-2\epsilon} ( \Delta \phi )  \; \theta(\Delta R   < R) 
\; 2 e^{-2 \Delta y   \epsilon } \, [(1-x) z]^{-1-2 \epsilon }\, (1-z)^{-1}
\nn \\
&\qquad  \quad \times 
\;  \frac{1- \min(z,1-z)^{2\epsilon}}{2\epsilon}
\,.
\end{align}
 
It is clear to see that the $1/\epsilon$ piece of this is identical to
\eq{naiveRsq}.
The $k_2^-$ soft gives an identical contribution. There remains the contribution
in which both $k_1^-$ and $k_2^-$ go soft simultaneously. Here the minus delta
function just becomes $\delta[(1-x)p^-]$ and factors out of the expression, and the remainder
looks very similar to a soft function calculation, so we can use the same
variables ($\Delta y, \Delta \phi, y_t, z$) as we use in that case. The
integration
for $y_t$ then looks like%
\footnote{Note that in the computation of $\Delta S^\two_{\rm base}$ and  $\Delta S^\two_{B,\mathrm{indep},1}$, see e.g. \eq{DeltaSB1uncorr2}, we obtain the same integral, except accompanied by an explicit theta function constraining $y_t>0$. Thus in the latter case we obtain $1/(4\epsilon)$ rather than zero.}
\begin{equation}
 \int_{-\infty}^{\infty} \df y_t \, e^{-4y_t\e }  = 0 \,.
\end{equation}
Thus, the net effect of the zero bin in the $1/\epsilon$ terms is to invert the
sign of \eq{naiveRsq}, to the sign predicted in \mycite{Tackmann:2012bt}.

We can also compute the full expression for the integral of the $\theta(\Delta R   < R) \times \mathcal{A}_A$ and zero-bin terms, including the finite pieces. 
Here we give the analytic expressions for the $\theta(\Delta R   < R) \times \mathcal{A}_A$ term:
\begin{align} 
\Bigl(\frac{\mu^2}{t^{\cut}}\Bigr)^{2\epsilon} & \Bigl( \frac{\alpha_s}{\pi} \Bigr)^2 \pi^2 C_i  \bigl[ \hat{P}^{(0)}_{ij}(x)(1-x) \bigr]
(1-x)^{-1-2\epsilon}
 \label{eq:thetaRanalytic} \\
& \times R^2 \biggl[ -\frac{1}{24} - \epsilon \biggl(\frac{1}{24}+\frac{9 \zeta_3}{8 \pi ^2}-\frac{\ln R}{12} +\frac{R^2}{288} +\frac{R^4}{17280} +\frac{R^6}{435456}\biggr) \biggr] + \mathcal{O}(R^{10}, \epsilon) \,, \nn
\end{align}
and the sum of the three zero bin terms,
\begin{align}  \label{eq:zerobinanalytic} 
\Bigl(\frac{\mu^2}{t^{\cut}}\Bigr)^{2\epsilon} & \Bigl( \frac{\alpha_s}{\pi} \Bigr)^2 \pi^2 C_i  \bigl[ \hat{P}^{(0)}_{ij}(x)(1-x) \bigr]
(1-x)^{-1-2\epsilon}
\\\nn
& \times R^2 \biggl[ -\frac{1}{12} - \epsilon \biggl( \frac{1}{12}+\frac{\zeta_3}{ \pi ^2}-\frac{\ln R}{6} +\frac{R^2}{144} +\frac{R^4}{8640} +\frac{R^6}{217728} \biggr) \biggr] + \mathcal{O}(R^{10}, \epsilon)
\,,\end{align}
which must be subtracted from the result of the naive beam function computation.

\section{\boldmath Beam function amplitudes in the small-\texorpdfstring{$R$}{R} limit}
\label{sec:smallRformulae}

\begin{figure}
\begin{center}
 \includegraphics[width=0.8\textwidth]{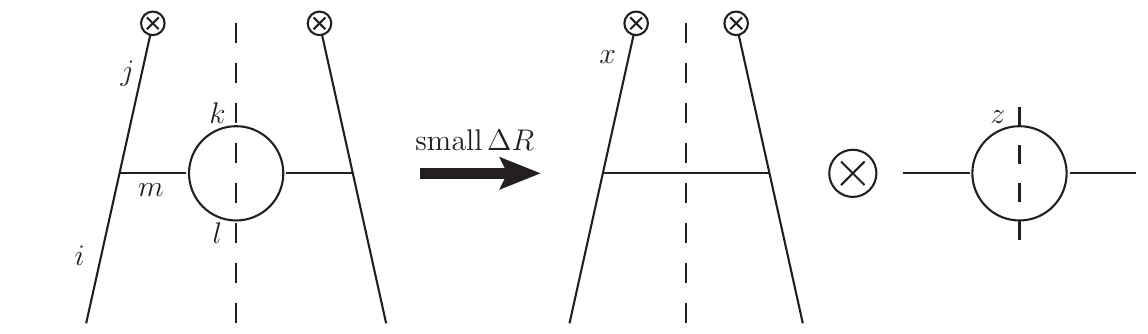}
 \caption{\label{fig:smallRgraph} The sole graph topology that, in light-cone gauge, corresponds to a divergent amplitude at small $\Delta R$. For small $\Delta R$, this decomposes into two splitting processes, as shown on the right hand side of the figure.
 }
\end{center}
\end{figure}

It is possible to write compact formulae for the small-$R$ limit of the beam function amplitudes $\mathcal{A}$ in terms of one-loop splitting functions. This is expected since at small $R$ the amplitude factorizes into two splitting
processes. The first is a splitting $i \to j + m$, where $i$ is the initial
state parton in $\mathcal{A}$, and $j$ is the physical parton that goes into 
the operator with momentum fraction $x$. The second is the splitting $m \to
k + l$, where $k$ and $l$ are the partons that go into the final state and 
generate the $\Delta R$. $m$ is an intermediate parton whose nature 
can be determined by quark number conservation from the other partons. 
As discussed in the main text, there is only one diagram that contributes
in the small-$R$ limit -- namely, the one in figure \ref{fig:smallRgraph}
that already has the topology of a diagonal $i \to j + (m \to k + l)$ process. Figure \ref{fig:smallRgraph} also illustrates the two splitting processes that this graph decomposes into at small $R$.

There are two distinct cases, corresponding to whether the intermediate parton 
$m$ is a gluon or a quark. We must treat the two cases differently because 
if one has a quark in the initial state, the plane of the splitting is not 
correlated to whether the quark has positive or negative helicity. However, 
when one has a gluon, there is a well-known correlation between the plane of
splitting and the gluon polarization.

The formula for the case when $m = q$ (or $\bar{q}$) is
\begin{align} \label{eq:smallRintmedQ}
 \mathcal{A}_R =& \hat{P}^{(0)}_{i\to jq}(x,\epsilon) \hat{P}^{(0)}_{q \to kl}(z,\epsilon) \frac{2g^4}{\Delta
R  ^2(p^-\Tau_T)^2(1-x)z(1-z)} 
\,,\end{align}
where $\hat{P}^{(0)}_{i\to jq}(x,\epsilon)$ originates from the first graph on the right hand side of figure \ref{fig:smallRgraph}, whilst $\hat{P}^{(0)}_{q \to kl}(z,\epsilon)$ originates in the second graph. The final factor in \eq{smallRintmedQ} arises from converting the transverse momentum denominators (etc.) in the splitting amplitudes
relative to the initial-state parton direction (for the first amplitude on the right hand side of figure \ref{fig:smallRgraph}), or relative to the parton $m$ (for the second amplitude), to our variables. In this formula one must 
always sum over all possibilities for $(k,l)$ -- i.e. $(q,g)$ and $(g,q)$. To get 
the full expression for $\mathcal{A}_R$ in $d$ dimensions, one must use 
the $d$-dimensional splitting functions given, e.g.\ in \mycite{Catani:1998nv}
\begin{align}
\hat{P}^{(0)}_{q \to gq}(x, \epsilon) &= C_F\biggl[ \frac{1+(1-x)^2}{x} - \epsilon x\biggr]
\,,\\
\hat{P}^{(0)}_{g \to q\bar{q}} (x, \epsilon) &= T_F \biggl[ 1 - \frac{2x(1-x)}{1-\epsilon}\biggr]
\,.\end{align}

The result when $m$ is an intermediate gluon is a little more complex and given by
\begin{align} \label{eq:ARgluonformula}
\mathcal{A}_R
&= \Bigl\{ \hat{P}^{(0)}_{i\to jg(\inpol)}(x, \epsilon) \left[ (\hat{P}^{(0)}_{g(\inpol) \to kl}(z)
\Delta y  ^2 + \hat{P}^{(0)}_{g(\outpol) \to kl}(z) \Delta \phi  ^2 \right]
\nn \\ & \quad
+ \hat{P}^{(0)}_{i\to jg(\outpol)}(x, \epsilon) \left[ (\hat{P}^{(0)}_{g(\outpol) \to kl}(z) \Delta y^2 + \hat{P}^{(0)}_{g(\inpol) \to kl}(z) \Delta \phi^2 \right]
\\\nn & \quad
 - 2\epsilon \hat{P}^{(0)}_{i\to jg(\outpol)}(x, \epsilon) \left[ (\hat{P}^{(0)}_{g(\outpol) \to kl}(z) (\Delta
y  ^2 + \Delta \phi  ^2) \right] \Bigr\}
\frac{2g^4}{\Delta R  ^4(p^-\Tau_T)^2(1-x)z(1-z)}
\,.\end{align}
The splitting functions $\hat{P}^{(0)}_{i \to jg(\inpol/\outpol)}(x, \epsilon)$ are splitting functions depending on
whether the final state gluon is polarized in or out of the splitting plane. These
functions may be computed from results in \mycite{Ellis:1991qj}. The two nonzero cases are given by
\begin{align}
\hat{P}^{(0)}_{q \to qg(\inpol)}(x, \epsilon) &= \frac{1}{2}C_F\frac{(1+x)^2}{1-x}
\,, \nn \\
\hat{P}^{(0)}_{q \to qg(\outpol)}(x, \epsilon) &= \frac{1}{2}C_F (1-x)
\,, \nn \\
\hat{P}^{(0)}_{g \to gg(\inpol)}(x, \epsilon) &= \frac{2C_A}{2-2\epsilon} \biggl[ (2-2\epsilon)\frac{x}{1-x} +
\frac{1-x}{x} + x(1-x) \biggr]
\,, \nn \\
\hat{P}^{(0)}_{g \to gg(\outpol)}(x, \epsilon) &= \frac{2C_A}{2-2\epsilon} \biggl[ x(1-x) + \frac{1-x}{x}\biggr]
\,.\end{align}
Here, the second outgoing parton is polarized, while the remaining spin/polarization indices are summed over or
averaged as appropriate, and $x$ is the momentum fraction of the first outgoing parton relative to the incoming quark.

Similarly $\hat{P}^{(0)}_{g(\inpol/\outpol) \to jk}(x, \epsilon)$ are splitting functions depending on the
polarization of the initial state gluon with respect to the splitting plane. These are given by
\begin{align}
\hat{P}^{(0)}_{g(\inpol) \to q\bar{q}}(x, \epsilon) &= n_f T_F (1-2x)^2
\,, \nn \\
\hat{P}^{(0)}_{g(\outpol) \to q\bar{q}}(x, \epsilon) &= n_f T_F
\,, \nn \\
\hat{P}^{(0)}_{g(\inpol) \to gg}(x, \epsilon) &= 2C_A \biggl[ \frac{1-x}{x} + \frac{x}{1-x} + (2-2\epsilon) x(1-x) \biggr]
\,, \nn \\
\hat{P}^{(0)}_{g(\outpol) \to gg}(x, \epsilon) &= 2C_A \biggl[ \frac{1-x}{x} + \frac{x}{1-x} \biggr]
\,,\end{align}
where $x$ is again the momentum fraction of the first outgoing parton relative to the incoming gluon.
In the case where the intermediate gluon decays into quarks one must remember to sum over the possibilities
$(j,k) = (q,\bar{q})$ and $(j,k) = (\bar{q},q)$.

Say, without loss of generality, that the initial parton travels along the $z$ direction,
and gluon $m$ is emitted somewhere in the $x-z$ plane. Then the first term corresponds 
to the case in which the gluon polarization lies in the $x-z$ plane (call this plane $P_1$).
If the gluon splits in plane $P_1$, resulting in $jk$ having a separation in $\Delta y  $, then 
the gluon polarization and splitting planes are coincident, so we should weight the $\Delta y^2$
factor with the $g(\inpol) \to kl$ splitting function. On the other hand, if the
gluon  splitting is in the plane which contains the  $m$ direction and the $y$
direction (call this plane $P_2$), the gluon polarization and splitting planes
are perpendicular. Here, partons $jk$ gain a separation in $\Delta \phi  $, so we weight the
$\Delta \phi  ^2$  factor with the $g(\outpol) \to kl$ splitting function. The second term in
\eq{ARgluonformula} corresponds to the case where gluon $m$ has polarization in the plane $P_2$.
Here we must swap the weighting factors multiplied by $\Delta \phi$ and $\Delta y$
around, for obvious reasons. The final term is actually where the gluon polarization lies in one of
the `extra' $-2\epsilon$ dimensions. Here, the gluon polarization is outside both splitting
planes. So both splitting function weightings are for the `out' polarization.

\section{Fit function coefficients for beam function}
 \label{sec:BFfitcoff}

\begin{landscape}

In these tables, the notation used for the numbers is such that $a^{-b}$ means $a \times 10^{-b}$ -- e.g. $4.197602^{-3}$ means $4.197602\times 10^{-3}$. The functional form for the fit is given in \eq{fitform_h}.

\noindent Coefficients for the low-$x$ fitting functions:
\vspace{2ex}

{\footnotesize{
 
\setlength{\tabcolsep}{0.25em}
\hspace{-7mm}\begin{tabular}{c || c | c | c | c | c | c | c | c | c | c | c}
Function & $1$ & $1/x$ & $x$ & $x^2$& $x^3$& $x^4$& $x^5$& $x^6$& $\ln x$& $\ln^2x$& $\ln^3x$\\
\hline
$gg$ $C_AC_A$ & 0.5599547 & -0.1723884 & -0.6712057 & 0.3811183 & 0.1042039 & -0.369331 & 0.2267327 & -5.925837$^{-2}$ & 4.197602$^{-3}$ & -5.069252$^{-2}$ & -3.139132$^{-4}$ \\
$gg$ $C_AT_F$ & 3.664257$^{-2}$ & -1.768324$^{-2}$ & -3.706551$^{-2}$ & 3.124682$^{-2}$ & -1.281309$^{-2}$ & 2.629221$^{-3}$ & -5.384836$^{-3}$ & 2.502594$^{-3}$ & -4.924704$^{-3}$ & -1.37947$^{-4}$ & -1.541345$^{-5}$ \\
$gg$ $C_FT_F$ & -2.459967$^{-2}$ & 2.760309$^{-2}$ & -1.78493$^{-2}$ & 7.339209$^{-2}$ & -0.1435441 & 0.166932 & -0.1188422 & 3.744709$^{-2}$ & 2.470372$^{-2}$ & 1.487655$^{-4}$ & 4.101906$^{-5}$ \\
$gq$ $C_AC_F$ & 4.539637$^{-2}$ & -0.1107837 & 6.281464$^{-2}$ & -0.1217796 & 0.2619967 & -0.35394 & 0.2722555 & -8.956735$^{-2}$ & -6.732607$^{-2}$ & -5.729447$^{-2}$ & -4.80465$^{-4}$ \\
$gq$ $C_FC_F$ & 6.919955$^{-2}$ &  -5.662405$^{-2}$ & -1.456111$^{-2}$ & -8.758454$^{-2}$ & 0.1992348 & -0.3054277  & 0.2616653 & -9.42774$^{-2}$ & 2.566124$^{-2}$ & 3.151035$^{-4}$ & 3.671834$^{-5}$ \\
$qg$ $C_AT_F$ & 9.072712$^{-2}$ & 0 & -0.1388956 & 0.2187705 & -0.3732872 & 
 0.4943961 & -0.3941701 & 0.1378447 & 5.811365$^{-2}$ & 2.3977$^{-2}$ & -6.644348$^{-5}$ \\
$qg$ $C_FT_F$ & -0.1293164& 0 & 0.2785842 & -0.2695549 & 8.224176$^{-2}$ & -0.1187513 & 0.1041828 & -3.930591$^{-2}$ & -1.120385$^{-2}$ & 1.534322$^{-4}$ & 7.011274$^{-6}$ \\
$qq$ $C_FC_A$ & 6.849965$^{-2}$ & 0 & 1.085542$^{-3}$ & -0.1488092 & 0.272654 & -0.4600145 & 0.4276455 & -0.1660631 & 1.282897$^{-3}$ & 6.464682$^{-4}$ & 3.027088$^{-5}$ \\
$qq$ $C_FC_F$ & -4.134826$^{-2}$ &0& 1.060512$^{-2}$ & 0.1243676 & -0.3495246 & 0.5930372 & -0.5528711  & 0.2147459 & -3.199847$^{-3}$ & -7.964259$^{-4}$ & -3.730933$^{-5}$ \\
$qq$ $C_FT_F$ & -1.040895$^{-3}$ &0& 8.209314$^{-3}$ & -3.441442$^{-3}$ & -7.79746$^{-3}$ & 7.739879$^{-3}$ & -5.269974$^{-3}$ & 1.619299$^{-3}$ & -2.244597$^{-3}$ & 8.752275$^{-7}$ & 1.688465$^{-7}$ \\
$q\bar{q}V$ & -3.638032$^{-2}$ &0& 0.1200994 & -0.2466278 & 0.3927571 & -0.4318602 & 0.2844866 & -8.334013$^{-2}$ & -1.104343$^{-2}$ & 1.492855$^{-3}$ & 7.829944$^{-5}$ \\
$qqS$ & 9.023778$^{-2}$ &0& -0.1345972 & 9.368575$^{-2}$ & -9.474317$^{-2}$ & 7.690635$^{-2}$ & -4.296148$^{-2}$ & 1.160153$^{-2}$ & 6.4119$^{-2}$ & 2.481566$^{-2}$ & -2.655248$^{-5}$ \\
\end{tabular}
}}

\vspace{2ex}
\noindent Coefficients for the high-$x$ fitting functions. Note that only the $q\bar{q}V$ fitting function has $(1-x)^8$ and $(1-x)^9$ terms.
\vspace{2ex}

 \footnotesize{
\setlength{\tabcolsep}{0.4em}
\hspace{-7mm}\begin{tabular}{c || c | c | c | c | c | c | c | c| c | c}
Function & $1$ & $(1-x)$ & $(1-x)^2$ & $(1-x)^3$ & $(1-x)^4$ & $(1-x)^5$ & $(1-x)^6$ & $(1-x)^7$ & $(1-x)^8$ & $(1-x)^9$\\
\hline
$gg$ $C_AC_A$ & 0 & 0.1160691 & -0.349332 & -6.242519$^-3$ & -0.1525609 & -0.5511262 & 0.689571 & -1.288973 & & \\
$gg$ $C_AT_F$ & 0 & 2.508153$^{-3}$ & -2.716604$^{-2}$ & -1.697566$^{-4}$ & -1.466446$^{-2}$ & -3.151329$^{-2}$ & 3.047598$^{-2}$ & -8.519766$^{-2}$ & & \\
$gg$ $C_FT_F$ & 0 & 1.297165$^{-2}$ & -1.023524$^{-4}$ & 2.363982$^{-2}$ & 1.84814$^{-2}$ & 5.227255$^{-2}$ & -6.220966$^{-2}$ & 0.1386245 & & \\
$gq$ $C_AC_F$ &-3.224531$^{-2}$ & -7.351504$^{-2}$ & -0.1300872 & -0.1554303 & -7.681572$^{-2}$ & -0.4687853 & 0.6436783 & -0.9973701 & & \\
$gq$ $C_F C_F$ & -2.630496$^{-2}$ & -3.540081$^{-2}$ & -7.299122$^{-2}$ & -6.800457$^{-2}$ & -4.575958$^{-2}$ & -0.1561224 & 0.1961347 & -0.3719774 & & \\
$qg$ $C_AT_F$ & 3.22456$^{-2}$ & -3.465512$^{-2}$ & 1.608474$^{-2}$ & 8.866111$^{-3}$ & 6.322399$^{-3}$ & 3.160005$^{-2}$ & -3.94882$^{-2}$ & 6.297286$^{-2}$ & & \\
$qg$ $C_FT_F$ & -9.079619$^{-2}$ & 0.202337 & -0.2195345 & 1.591669$^{-3}$ & 2.171154$^{-3}$ & 1.184325$^{-3}$ & 2.465939$^{-3}$ & 3.120635$^{-3}$ & & \\
$qq$ $C_FC_A$ & 0 & 0.1196236 & -5.012343$^{-2}$ & 7.405496$^{-4}$ & -2.196023$^{-4}$ & -3.252667$^{-6}$ & -5.979243$^{-4}$ & 2.384104$^{-4}$ & & \\
$qq$ $C_FC_F$ & -7.410597$^{-3}$ & -3.199028$^{-2}$ & -1.492514$^{-3}$ & 4.92313$^{-4}$  & 3.723146$^{-4}$ & 1.562987$^{-3}$ & -1.72615$^{-3}$ & 2.269269$^{-3}$ & & \\
$qq$ $C_FT_F$ & 0 & 1.022712$^{-2}$ & -8.963093$^{-3}$ & 1.484597$^{-3}$ & 4.828097$^{-4}$ & 1.733115$^{-3}$ & -2.325544$^{-3}$ & 2.940276$^{-3}$ & & \\
$q\bar{q}V$ & 0 & 0 & 0 & 3.200958$^{-5}$ & 3.236335$^{-5}$ & 4.271679$^{-4}$ & 6.238066$^{-4}$ & 2.188824$^{-3}$ & -2.43424$^{-3}$ & 6.561077$^{-3}$ \\
$qqS$ & 0 & 6.485799$^{-3}$ & 8.372154$^{-5}$ & 9.840991$^{-3}$ & 8.189228$^{-3}$ & 1.847319$^{-2}$ & -1.837079$^{-2}$ & 4.499479$^{-2}$ & & \\
\end{tabular}
}

\end{landscape}

\phantomsection
\addcontentsline{toc}{section}{References}
\bibliographystyle{../jhep}
\bibliography{../beamfunc}

\end{document}